    \documentclass[review]{elsarticle}
    \usepackage{lineno}
    \usepackage{graphics}
    \usepackage{graphicx}
    \usepackage[caption=false]{subfig}
    \usepackage[utf8]{inputenc}
    \usepackage{ifthen}
    \usepackage{algorithm}
    \usepackage{algorithmic}
    \usepackage{a4wide}
    \usepackage{amssymb,amsbsy,amsmath,mathtools}
    \usepackage{empheq}
    \usepackage{hyperref}
    \usepackage{xargs}
    \usepackage{xcolor}
    \usepackage[colorinlistoftodos,prependcaption,textsize=tiny]{todonotes}
    \usepackage{soul}
    
    \usepackage{tikz}
    	\usetikzlibrary{calc}
    	\usetikzlibrary{shapes,shadows,arrows,positioning}
    	\usetikzlibrary{decorations.pathreplacing,decorations.pathmorphing}
    
    \usepackage{pgfplots}
    	\usepgfplotslibrary{colorbrewer}
    	\usepgfplotslibrary{groupplots}
        \pgfplotsset{compat=newest}
        
    \pgfplotsset{height=0.4\linewidth,}
    \pgfplotsset{    
    		x label style={at={(axis description cs:0.5,-0.1)},anchor=north},
        	y label style={at={(axis description cs:-0.1,.5)},anchor=south},
            ylabel near ticks,
    		ylabel shift = -0.5em,
    }
    \pgfplotsset{every axis/.append style={
                        legend style={font=\tiny,line width=.5pt,mark size=.6pt},
                        }}
    \pgfplotsset{
    		axis on top,
            xtick align=inside,
    }
    \pgfplotscreateplotcyclelist{newcolors}
    {%
    blue,           every mark/.append style={fill=blue!80!black},mark=*\\%
    red,            every mark/.append style={fill=red!80!black},mark=square*\\%
    brown!60!black, every mark/.append style={fill=brown!80!black},mark=otimes*\\%
    black,          mark=star\\%
    blue,           every mark/.append style={fill=blue!80!black},mark=diamond*\\%
    red,densely dashed,every mark/.append style={solid,fill=red!80!black},mark=*\\%
    brown!60!black,densely dashed,every mark/.append style={
    solid,fill=brown!80!black},mark=square*\\%
    black,densely dashed,every mark/.append style={solid,fill=gray},mark=otimes*\\%
    blue,densely dashed,mark=star,every mark/.append style=solid\\%
    red,densely dashed,every mark/.append style={solid,fill=red!80!black},mark=diamond*\\%
    }
    
    \pgfplotsset{
      log x ticks with fixed point/.style={
          xticklabel={
            \pgfkeys{/pgf/fpu=true}
            \pgfmathparse{exp(\tick)}%
            \pgfmathprintnumber[fixed relative, precision=3]{\pgfmathresult}
            \pgfkeys{/pgf/fpu=false}
          }
      },
      log y ticks with fixed point/.style={
          yticklabel={
            \pgfkeys{/pgf/fpu=true}
            \pgfmathparse{exp(\tick)}%
            \pgfmathprintnumber[fixed relative, precision=3]{\pgfmathresult}
            \pgfkeys{/pgf/fpu=false}
          }
      }
    }

    \tikzset{
    >=latex,
    punkt/.style={
           rectangle,
           rounded corners,
           draw=black, very thick,
           text width=6.5em,
           minimum height=2em,
           text centered}
}

    \newcommand{\LVM}{Lotka-Volterra model~}
    \newcommand{\ZEVs}{zero-emissions vehicles~}
    
    \newcommand{\HFCVs}{hydrogen fuel cell vehicles~}
    \newcommand{\HBVs}{hydrogen-based vehicles}
    \newcommand{\prey}{N_v} % x
    \newcommand{\predator}{N_h} %y
    \newcommand{\grate}{\gamma}

    \newcommand{\gprey}{{\grate}_c}
    \newcommand{\gpredator}{{\grate}_h}
    \newcommand{\attack}{a}
    \newcommand{\efficiency}{\epsilon}
    \newcommand{\Nvehicle}{{N_v}}
    \newcommand{\massv}{m_i}
    \newcommand{\dmassv}{\dot{m}_i}
    
    \newcommand{\masstot}{m}

\begin{document}

    \newcommandx{\unsure}[2][1=]{\todo[linecolor=red,backgroundcolor=red!25,bordercolor=red,#1]{#2}}
    \newcommandx{\change}[2][1=]{\todo[linecolor=blue,backgroundcolor=blue!25,bordercolor=blue,#1]{#2}}
    \newcommandx{\info}[2][1=]{\todo[linecolor=black!50!green,backgroundcolor=green!25,bordercolor=black!50!green,#1]{#2}}
    \newcommandx{\improvement}[2][1=]{\todo[linecolor=gray,backgroundcolor=gray!25,bordercolor=gray,#1]{#2}}

\newcommand{\franky}[1]{ {\color{red}{FL: #1 }} }

% %    \titlerunning{Forecast of Hydrogen Vehicle fleet.}   
%     \author{Sahdia Maryam \& Florimond Gu{\'e}niat}
%     %\authorrunning{Short form of author list} % if too long for running head
%  %   \institute{
% %    S. Maryam \& F. Gueniat at
%  \address{
%     	Birmingham City University, Department of Mechanical Engineering\\
%         \email{florimond.gueniat@bcu.ac.uk}            
%     }

%\maketitle
\begin{frontmatter}
\title{Forecasting the Evolution of Hydrogen Vehicle Fleet in the UK using Growth and Lotka-Volterra Models.}
\author[bcu]{Sahdia Maryam }
\author[bcu]{Florimond Gu{\'e}niat\corref{cor1}}

\address[bcu]{Birmingham City University, Department of Mechanical Engineering, UK }

\cortext[cor1]{corresponding author. \url{florimond.gueniat@bcu.ac.uk}}

\begin{abstract}
    Road vehicles play an important role in the UK’s energy systems and are a critical component in reducing the reliance on fossil fuels and mitigating emissions. 
    A dynamic model of vehicle fleet, based on predator-prey concepts, is presented. 
    It allows to predict the evolution of the hydrogen based vehicle’s role in the UK’s vehicle fleet and the sensitivity of this growth to the supply chain. 
    In addition to this, this model is used to predict the demand of hydrogen for the passenger vehicle fleet for various scenarios. 
    A key result is that the amount of hydrogen required to support a huge network of hydrogen based vehicles is currently not practical, regarding the infrastructure required and the cost to build such an infrastructure.
    In order to mitigate that, the policy focus should move primarily from hydrogen based vehicles to first encompass range extenders in the transport energy system.

\end{abstract}

\end{frontmatter}

\section*{Highlights}
\begin{itemize}
    \item A comprehensive and predictive model of vehicle fleet growth.
    \item The model includes the fleet composition, in particular hydrogen based vehicles and conventional fuel based vehicles.
    %\item A quantification of the importance of the parameters of this model, which relates to fuel supply chain
    \item A study of the UK market and policy discussions with respect to policy objectives, such as the Road to Zero.
\end{itemize}

%%%%%%%%%%%%%%%%%%%%%%%%%%%%%%%%%%%%%%%%%%%%%%%%%%%%%%%%%%%%%%%%%%%%
%%%%%%%%%%%%%%%%%%%%%%%%%% SECTION %%%%%%%%%%%%%%%%%%%%%%%%%%%%%%%%%
%\begin{itemize}
%\item In the section 3.3. it starts Three scenario have been designed: Try to provide some introduction as it starts really out of the blue with scenarios. I think it would be better to introduce the policies fist and then say the scenarios 
%\item I think the section 3.3 and 4 need a little more elaboration/clarification
%\item Moreover, it would be nice more discussion of the findings (bring other studies, compare them)
%\item Discuss also some further contribution e.g. policy, and future research opportunities 
%\end{itemize}

\begin{figure}[ht]
\center
\begin{tikzpicture}[node distance=3cm, auto,]
 %nodes
 \node[punkt] (cfleet) {conventional fleet};
 
 \node[punkt, above left of=cfleet]
 (crate) {consumption of ressources};

 \node[right of=cfleet,xshift=-1cm]
 (cplus) {};

 \node[punkt,right of=cplus,xshift=-1cm]
 (csupply) {supplies (fuel)};

 \node[right of=csupply,xshift=-1cm]
 (cequal) {};
 
 \node[punkt,right of=cequal,xshift=-1cm]
 (cchange) {change in conventional fleet};
 
\draw[very thick] ($(cplus) + (-0.4,0)$) -- ++(0.8,0);
\draw[very thick] ($(cplus) + (0,-0.4)$) -- ++(0,0.8);

\draw[thick] ($(cequal) + (-0.25,-0.1)$) -- ++(0.5,0);
\draw[thick] ($(cequal) + (-0.25,+0.1)$) -- ++(0.5,0);

\node (cminus) at ($(crate)!0.5!(cfleet)$) {};

\draw[very thick] ($(cminus) + (-0.25,0)$) -- ++(0.5,0);

\draw[->, line width=1mm, blue] (crate.east) to [out=0,in=90] (cfleet.north);
\draw[->, line width=1mm, blue] (cfleet.west) to [out=180,in=-90] (crate.south) ;

 \node[punkt,below of = cfleet,yshift=-0cm] (hfleet) {hydrogen fleet};
 
 \node[punkt, below left of=hfleet]
 (hrate) {consumption of ressources};

 \node[right of=hfleet,xshift=-1cm]
 (hplus) {};

 \node[punkt,right of=hplus,xshift=-1cm]
 (hsupply) {supplies (hydrogen)};

 \node[right of=hsupply,xshift=-1cm]
 (hequal) {};
 
 \node[punkt,right of=hequal,xshift=-1cm]
 (hchange) {change in hydrogen fleet};
 
\draw[very thick] ($(hplus) + (-0.4,0)$) -- ++(0.8,0);
\draw[very thick] ($(hplus) + (0,-0.4)$) -- ++(0,0.8);

\draw[thick] ($(hequal) + (-0.25,-0.1)$) -- ++(0.5,0);
\draw[thick] ($(hequal) + (-0.25,+0.1)$) -- ++(0.5,0);

\node (hminus) at ($(hrate)!0.5!(hfleet)$) {};

\draw[very thick] ($(hminus) + (-0.25,0)$) -- ++(0.5,0);

\draw[->, line width=1mm, blue] (hrate.east) to [out=0,in=-90] (hfleet.south);
\draw[->, line width=1mm, blue] (hfleet.west) to [out=180,in=90] (hrate.north) ;

\node[punkt] (transition) at ($(cfleet)!0.5!(hfleet)$) {transition};

\draw[->, line width=1mm, blue] (cfleet.south east) to [out=-45,in=45] node[midway,above,sloped] {increase} (hfleet.north east) ;

\draw[->, line width=1mm, blue] (hfleet.north west) to [out=135,in=-135]  node[midway,above,sloped] {decrease} (cfleet.south west) ;

\end{tikzpicture}
\caption{Graphical Abstract: dynamic model of the growth of vehicle fleets}
\label{fig-abstract}
\end{figure}
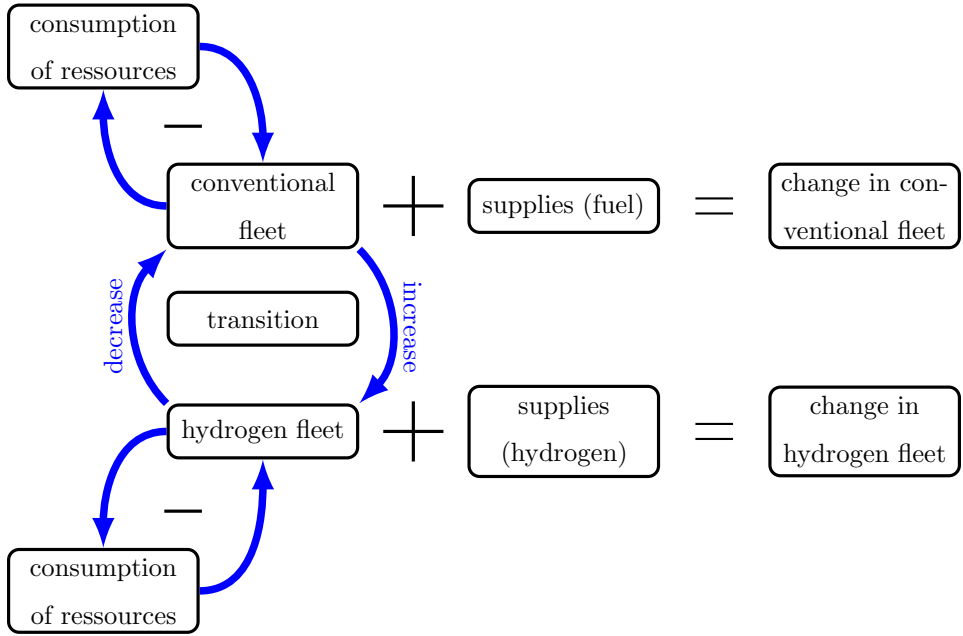

%\linenumbers
\section{Introduction} \label{sec-intro}
    
    The global energy consumption rate is increasing annually, even after the signature of the Paris Agreement. In particular, the demand for fossil fuels is still increasing, ~\cite{IEA2020}.
    The transportation sector is currently responsible for $27\%$ of energy demand,~\cite{Haslam2012}, thus impacting the level of greenhouse gases and pollution produced,~\cite{Acar2014}.
    Consequently, governing bodies are putting into place policies to transition to a low carbon transport system.
    There is indeed a major shift in the public policies ; almost twenty countries are planning on removing diesel and petrol cars from the roads,~\cite{Meckling2019}.
    In particular, the UK government has recently planned the "road to zero" policy. The UK's objectives are simple but ambitious: the entire fleet should be composed by zero-emissions vehicles by 2050,~\cite{Roadtozero}. By 2030, no more conventional vehicles will be sold.

    Such policies can drive the increasing demand and investments in alternative fuel vehicles,~\cite{Ito2013,Hackbarth2013}, to replace the internal combustion engine vehicles (ICEVs) fleet with zero-emissions vehicles. 
    Two technologies for \ZEVs seem promising at this point: 
    the celebrated electric vehicles, and hydrogen-based vehicles (HBVs).
    Using hydrogen in a fuel cell only produces water and heat as waste, eliminating emissions produced by internal combustion engine vehicles. 
    However, the production and distribution of hydrogen is associated with an energy cost, and so, emphasis on producing hydrogen from renewable energy resources is critical for long-term green, sustainable transport network,~\cite{Abbasi2011,Momirlan2005,TolgaBalta2009,Fayaz2012,Bolat2014}. 
    Noteworthy, hydrogen is not only a possible fuel, but can be used as well for developing energy storage systems,~\cite{Momirlan2005,Dincer2015,Liu2012}, effectively mitigating some issues related to renewable energy resources. 
    Consequently, hydrogen has the potential to play a fundamental role in tomorrow’s energy system,~\cite{Fayaz2012,Dagdougui2012,Edwards2008,Neef2009,Mazloomi2012,Pudukudy2014,Sharma2015}. 
    The main obstacle hindering vehicle manufacturers and consumers from embracing hydrogen fuel cell vehicles are the high cost, high levels of complexity associated with the hydrogen supply chain infrastructure, i.e. in terms of the options available at each infrastructure node and recharge stations,~\cite{Fayaz2012,Bolat2014,Dagdougui2012,Almansoori2006}. 
    Constructing new infrastructures is indeed extremely costly, and poor choice in the design of the support infrastructure can lead to failure in the transition,~\cite{Stephan2004}.
    As a matter of fact, refuelling stations in London deliver hydrogen from electrolysis at around \pounds 0.20/kWh, making refuelling twice more expensive than petrol.
    To this date, there is only a dozen of hydrogen refuelling station in the UK.
    Indeed, one of the key challenges is to secure sufficient investments when there is no assurance of profitable demand, which is enslaved to the \HBVs fleet. 
    Moreover, it has been shown that a growth of the delivery infrastructure triggers a feedback mechanism, sustaining in return the growth of the hydrogen supply chain and of the \HBVs,~\cite{Haslam2012,Dagdougui2012}.
    It means that governing policies have a decisive role to play in encouraging the transition,~\cite{Brand2020,Haslam2012}.
    Strong policies are also put into place in EU, with the Hydrogen Strategy,~\cite{EU2020}, and Japan,~\cite{METI2019}.
    
    %The main aim of this paper is to deliver a realistic model of the evolution of the HFCVs fleet. 
    To understand the current and future environmental impact of the UK road transportation system, models able to represent fleet evolution are needed.
    Such models are pivotal to develop policies through developing and evaluating scenarios and their consequences,~\cite{Kloess2011}.
    To the author's knowledge, models developed so far, as in~\cite{Stephan2004,Wang2011,Park2011}, do not encompass all the following properties: i) the general growth of the fleet ii) the coexistence and the nonlinear interactions/feedback between the different types of advanced vehicles in the fleet composition, iii) the role of the supply chain in energy.
    Considering the fleet size is fundamental to design the infrastructure, but is usually overlooked,~\cite{Stephan2004,Moreno2017},
    and predicting growth and decline allows to re-purpose the infrastructure,~\cite{Samsatli2019}.
    Fleet composition is mostly tackled as a transitional process based on external factors (such as energy prices or the supply chain infrastructure, to name a few),~\cite{Stephan2004,Zhao2006,Shafiei2017} rather than a competition process including feedback loop and the fleet composition as a decisive factor,~\cite{Musti2011}. 
    %\todo{how much impact of environment in UK ? policy by 2050 ?}
    The UK intends to reduce its carbon account by at least 100\% for the year 2020 than the 1990 baseline,~\cite{act2019}. 
     The UK government announced in 2017 that it will end the sale of all new conventional petrol and diesel cars and vans by 2040,~\cite{act2017}.
     It has recently been announced that it will be moved forward to 2030. 
     It means drastic changes in the composition of vehicle fleet in the UK will happen, with associated changes and stress in the chain supply of fuel.
     
    The main aim of this paper is to deliver a realistic model of the evolution of the conventional and zero-emission fleets that can be used to efficiently inform policy makers. 
    It means that the model has to be comprehensible, informative and take into account the fleet composition, for policies to be crafted.
    Additionally, the model has to reflect reality and has to be able to predict the effects of policies.
    A dynamic systems based approached is chosen, for the flexibility and the zoology of models that already exists,~\cite{Morrison2012}. 
    The model is a modified Lotka-Volterra model (LVM). 
    The LVM, also known as predator-prey model, is based on the growth model exploring interactions between two or more diverse competitors,~\cite{Wang2016}. 
    The Lotka-Volterra equations have frequently been utilised in research areas modelling competing technologies,~\cite{Zhang2014}, despite being defined originally to analysis problems concerning population dynamics ,~\cite{Gokmen2015}.
    It has been modified to reflect the effects of the supply chain in energy.
    The model is validated in the context of the United Kingdom’s market. 
    The model delivers two main results. 
    Firstly, it allows to accurately predicting the evolution of the \HBVs~fleet. 
    As stated earlier, it is fundamental to secure investments and hence sustain the growth of the fleet, hence triggering a reinforcement loop. 
    Secondly, it allows to run predictive simulations with respect to different scenarios, and hence to determine the needs or efficiency of policies.
    %A decisive need for the development of HBVs is to reach a critical amount of funding. 
    %The aim is to identify the strategy which will maximise the chances of \HBVs penetrating the market and growing to a mass production level,~\cite{Zhang2014}. 
    %Running the model against various scenarios allows determining which strategies vehicle manufacturers should take alongside other stakeholders. 
    
    The paper is organized as follows. The framework and model are discussed in Sec.~\ref{sec-model}.
    The resulting model is validated and its mathematical properties are investigated, in Sec.~\ref{sec-res}.
    Efficient policies are discussed, in the case of UK, Sec.~\ref{sec-policies}.
    Concluding remarks close the paper in Sec.~\ref{sec-outro}.

%The results and discussion is split into sections following the order of objectives of this paper. 
            %Firstly, the results depicting the growth of conventional vehicles is given and model validation in Sec.~\ref{ssec-growth_val}. 
            %Then the model is extended to encompass predator-prey aspects giving the second order LVM and validation of the model, Sec.~\ref{ssec-lvm_val}. 
 %           Sec.~\ref{ssec-policies} addressing the so called hydrogen "space" available for the transition to occur is provided before a comparison of \HFCV and HFC-REs impact in introducing hydrogen as a transportation fuel.       

%%%%%%%%%%%%%%%%%%%%%%%%%%%%%%%%%%%%%%%%%%%%%%%%%%%%%%%%%%%%%%%%%%%%
%%%%%%%%%%%%%%%%%%%%%%%%%% SECTION %%%%%%%%%%%%%%%%%%%%%%%%%%%%%%%%%

\section{Model} \label{sec-model}  
    Lack of historical data of \HFCVs can be a limitation in using forecasting models. 
    Developing economic growth model driven by data from conventional vehicles may overcome this issue. 
    Additionally, utilising dynamical systems to model the hydrogen supply chain opens up many possibilities in integrating various powerful mathematical tools. 
    %Product Diffusion Model
    
    %(PDM) Forecasts the market penetration of new products by considering the lifecycle. (Wang and Wang, 2016a)
    The product diffusion model~\cite{Wang2016} is not able to fully explain the diffusion behaviour and mechanism in a mutually competitive market. %lack of interaction between models 
    Since, non fuel cell hybrids, electric vehicles and other ultra-low emissions vehicles are also expected to compete in the road transport market making the representation of mutual interaction necessary, for this reason the Lotka-Volterra model (LVM) is considered and presented Sec.~\ref{ssec-lvm}.
    The classical predator-prey system was first considered by Lotka in 1920 modelling undamped oscillations for chemical reactions and then later by Volterra to predator-prey interactions,~\cite{Evans1999}. 
    %One of the features of the \LVM is the oversimplification, where it lacks robustness and often demonstrates mathematical instability against various model modifications and variations [56]. 
    
    These equations will be modified for reflecting more accurately the market in Sec.~\ref{ssec-modif}, using economic growth principles.
    Growth curve modelling is a generic term used in different contexts encompassing various statistical models to map the growth of a product. 
    %Indeed, the exponential equation is a standard equation that describes the growth of a single population [90]. 
    %However, more recently, this term has come to refer to a discrete set of analytical approaches. 
    %Typically the term growth curve model refers to statistical methods that enable estimation of inter-individual variability in intra-individual patterns of change over time. 
    %A growth model is an approach used to map the growth of things from microbiology to populations. 
    %This method has proved to be very popular within the social and statistical science
    
    %!!!!!!!!!!!!!!!!!!!!!!!!!! SUBSECTION !!!!!!!!!!!!!!!!!!!!!!!!!!
    \subsection{Lotka-Volterra Model}\label{ssec-lvm}
        The predator-prey interaction consists of a pair of first order autonomous ordinary differential equations as follows.

        The evolution of the number of preys is modelled by Eq.~\eqref{eq-prey}: 
       \begin{equation}
            \dot{\prey} = \prey(\gprey - \attack \predator)
            \label{eq-prey}
        \end{equation}
        The number of prey,$\prey$, increases with its own growth rate $\gprey$, and decreases with the attacks of predators. 
        This decrease depends on the number of predators $\predator$, the attack rate $\attack$ of the predators and the number of available prey.
        The units of the rate of attack is $time^{-1} predator^{-1}$. 
        It simply translates both the pressure on the population of preys and the effect on the growth rate, induced by the presence of the predators.
        
        The number of predator evolves similarly:
        \begin{equation}
            \dot{\predator} = \predator(\efficiency \prey - \gpredator)
            \label{eq-predator}
        \end{equation}
        
        In Eq.~\eqref{eq-predator}, prey are replaced by predators in the system with an efficiency $\efficiency$. 
        The units of the efficiency is $time^{-1} prey^{-1}$. 
        It simply translates that the growth rate is mostly driven by the available food supply (the preys).
        However, predators compete for the supply ; a negative sign is associated with the growth rate $\gpredator$.

        The \LVM model (LVM) is the coupling of Eqs.~\eqref{eq-prey} and~\eqref{eq-predator}:
        
        \begin{equation}
        \left\{
            \begin{array}{ccc}
            \dot{\prey} &=& \prey(\gprey - \attack \predator)\\
            \dot{\predator} &=& \predator(\efficiency \prey - \gpredator)
            \end{array}
        \right.
        \label{eq-lvm}
        \end{equation}

        \begin{figure}
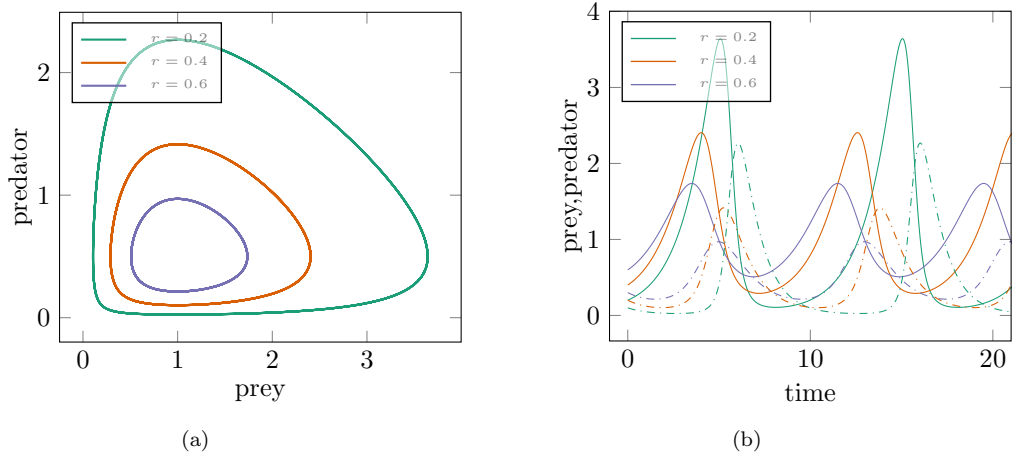

            \center
            \subfloat[]{\input{./tikz/tikz_preamble}

\pgfplotsset{cycle list/Dark2-6}
\begin{tikzpicture} 
    \begin{axis}[
        %clip mode=individual,
        cycle list name=newcolors,
        legend pos=outer north east,
		xlabel={prey},
		ylabel={predator},
        %ytick = {1.e-5,1.e-3,1.e-1},
        %xtick = {1.e-4,1.e-3,1.e-2,1.e-1,1},
        %xmin=1.e-2,xmax=5,
        %ymin=5.e-6,ymax=.2,
        %cycle multi list={cycle list=Set1-4}
        cycle list name=Dark2-6,
        legend pos=north west,
        legend style={fill=white, fill opacity=0.5, draw opacity=1,text opacity=1},
        ]
        \addplot+[thick] 
            table{./data/pp_phase_x_0.20_y_0.10.csv};

        \addplot+[thick]
            table{./data/pp_phase_x_0.40_y_0.20.csv};
        
        \addplot+[thick]
            table{./data/pp_phase_x_0.60_y_0.30.csv};
        
        %\addplot+[thick]
         %   table{./data/pp_phase_x_0.80_y_0.40.csv};
        
       % \addplot+[thick,mark=x]
        %    table{./data/pp_phase_x_1.00_y_0.50.csv};

        %\addplot+[thick]
        %    table{./data/pp_phase_x_1.20_y_0.60.csv};
        \legend{$r=0.2$,$r=0.4$,$r=0.6$}

    \end{axis}
\end{tikzpicture}\label{sfig-pp_phase}}
            \subfloat[]{\input{./tikz/tikz_preamble}

\pgfplotsset{cycle list/Dark2-3}
\begin{tikzpicture} 
    \begin{axis}[
        %clip mode=individual,
        cycle list name=newcolors,
        legend pos=outer north east,
		xlabel={time},
		ylabel={prey,predator},
        xtick = {0,10,20},
        xmin=-1,xmax=21,
        %ymin=5.e-6,ymax=.2,
        %cycle multi list={cycle list=Set1-4}
        legend pos=north west,
        legend style={fill=white, fill opacity=0.5, draw opacity=1,text opacity=1},
        cycle list name=Dark2-3,
        ]
        \addplot+[] 
            table{./data/pp_xdata_x_0.20_y_0.10.csv};

        \addplot+[]
            table{./data/pp_xdata_x_0.40_y_0.20.csv};
        
        \addplot+[]
            table{./data/pp_xdata_x_0.60_y_0.30.csv};

        \addplot+[dash dot] 
            table{./data/pp_ydata_x_0.20_y_0.10.csv};

        \addplot+[dash dot]
            table{./data/pp_ydata_x_0.40_y_0.20.csv};
        
        \addplot+[dash dot]
            table{./data/pp_ydata_x_0.60_y_0.30.csv};
        
            \legend{$r=0.2$,$r=0.4$,$r=0.6$}

    \end{axis}
\end{tikzpicture}\label{sfig-pp_ts}}
            \caption{Prey-predator system, with $\gprey=2/3$, $\attack=4/3$, $\efficiency=\gpredator=1$. Initial conditions are $(N,P)=(r,0.5r)$, for various $r$. (a): Phase plot. (b): Time series. Solid lines correspond to preys, dashed lines to predators.}
            \label{fig-pp}
        \end{figure}
        
        The LVM results in periodic growth and decay for both the predator and prey in response to the growth and decay of the other, see Figs.~\ref{fig-pp}.
        This is suitable and broadly explains the dynamic phenomena in population ecology,~\cite{Huang2001}. 
        However, such an oscillatory behavior is not expected nor realistic for the evolution of conventional and \HBVs~fleets.
        In the next two sections, the \LVM is modified to encompass more realistic dynamics.

    \subsection{Modified growth rate} \label{ssec-growth}
        %Developing the growth model was essential to modify the LVM, because one of the limitations of the LVM was its over-simplicity 
        The growth of conventional vehicles has been historically modelled as linear,~\cite{Leibling2008,Wei2015,Hugosson2016}. 
        It means the fleet is expected to grow without a cap.
        %It is both unrealistic and uncomon
        %
        It is intuitively incorrect. 
        The fleet growth is expected to eventually decline due to several factors i) the availability of fuel ii) the maturity of the market iii) the introduction of more desirable and sustainable alternatives.

        However, the transient growth can be seen, in a first approximation, as linear in first order models. The seemingly linear growth can be the sign of a non-mature market. 
        Based on this remark, %we introduce the following growth model.
        %,, to incorporate a more realistic growth model representing UK’s vehicle growth over the last 50 years. 
        a first order equation, based on the available energy, is introduced, to map the growth of conventional vehicles. % in the UK from 1961 to the present day. 
        The total mass of fuel $\dot{m}_t$ used by the fleet of vehicles, per year, is calculated by combining the rate of mass being consumed, $\dot{m}$, with the mass wasted $\dot{m}_w$: 
        \begin{equation}
            \dot{m}_t = \dot{\masstot}+\dot{m}_w
            \label{eq-total_mass}
        \end{equation}
        
        In the following, a simplifying assumption is made, by equating the mass wasted to zero, and hence $m_t=\masstot$.
        
        The total mass of fuel consumed depends on the number of vehicles $\Nvehicle$ times the mass of fuel consumed per vehicle, $\massv$:
        
        \begin{equation}
            \masstot=\Nvehicle \times \massv
            \label{eq-mass}
        \end{equation}
        
        Note that this equation can be applied to any fleet. For conventional IC vehicle, the total mass $\masstot$ would be the mass of petrol used per year, while for hydrogen vehicle, it would be the mass of hydrogen being used per year.
        \smallskip
        
        Differentiating Eq.~\eqref{eq-mass} leads to the rate of total mass consumed per year:
        
        \begin{equation}
            \dot{\masstot}=\dot{\Nvehicle} \times \massv + \Nvehicle \times \dmassv
            \label{eq-dmass}
        \end{equation}
        
        Rearranging:
        
        \begin{equation}
            \begin{array}{ccl}
                \dot{\Nvehicle} &=& -\dfrac{\dmassv }{\massv}\Nvehicle + \dfrac{\dot{\masstot}}{\massv}\\
                &=& -\grate\Nvehicle + \mu
            \end{array}
            \label{eq-grate}
        \end{equation}
        where we note $\grate := \dfrac{\dmassv }{\massv}$ (per years, or in $years^{-1}$) and $\mu := \dfrac{\dot{\masstot}}{\massv} $ (in vehicles per year, or $Mveh/year$).
        The quantity $\mu$ is the mass of fuel used annually divided by the mass of fuel consumed by a car. 
        It corresponds to the maximum number of new vehicles that can be sustained, on a yearly basis, by the supply chain of fuel.
        %In other words, it corresponds to the total available resources in fuel.
        
        The amount of fuel consumed, annually, by a car is positive and can be approximated by a constant (it holds as long as there is no breakthrough in technology).
        The quantity $\grate$ is the mass of fuel used annually by a car divided by the mass of fuel consumed by a car. 
        This is an inverse time scale, which corresponds to the time needed by the fleet to absorb all the resources.
        In other word, it is the rate at which the system evolves. 
        It corresponds to a growth rate.
        The rate, $\grate$ is actually positive.
        It has a powerful implication: the actual rate is driven by the resources $\mu$; without resources, the number of vehicles tend to zero. 
        The presence of the source term mitigates this dynamics: the available fuel, which acts as an available resource, drives up the number of vehicles.
        
        One can assume that the individual car consumption, $\dot{\massv}$, is constant, and that the fuel consumption, per year, will reach either an optimal or maximal value.
        
        Taking the limit, the number of vehicle will then reach a plateau as well:
        \begin{equation}
            \lim\limits_{t\rightarrow \infty} \Nvehicle = \dfrac{\dot{\masstot}}{\dot{\massv}}
            \label{eq-limit}
        \end{equation}
        This value depends exclusively on the total fuel resources available for consumption per year $\dot{\masstot}$, divided by the individual car consumption $\dmassv$.
        It means that $\lim\limits_{t\rightarrow \infty} \Nvehicle$ is the number of cars that can, ultimately, be supported by the system, \emph{i.e.}, by the supply chain of fuel.
        It is a fundamental difference with a linear growth model, that diverges. 

    %!!!!!!!!!!!!!!!!!!!!!!!!!! SUBSECTION !!!!!!!!!!!!!!!!!!!!!!!!!!
    \subsection{Modified Lokta-Volettra Model}\label{ssec-modif}
        The growth model from Eq.~\eqref{eq-grate} does not allow to represent competition between conventional and hydrogen fleets.
        However, it enlights on how to modify the LVM equations from Eqs.~\eqref{eq-prey} and~\eqref{eq-predator} to take into account the realistic development of the fleet.
        
        The growth rates are, in these equations, respectively $\gprey$ and $\gpredator$ for the preys and the predators.
        We note $\predator:=\Nvehicle_h$ the number of hydrogen vehicle, and $\prey:= \Nvehicle_c$ the number of conventional vehicle. 
        Entering the market, we consider the hydrogen vehicles as the predators and the conventional vehicles as the preys.
        
        Considering the remarks from Sec.~\ref{ssec-growth}, we propose a model more realistic for the competition between conventional and non conventional vehicles.
        
        The growth rate are for conventional fleet $\gprey$ is replaced, in Eq.~\eqref{eq-grate}, with the coupling term from Eq.~\eqref{eq-lvm}:
        
        \begin{equation}
            \gprey \leftarrow -\gprey - \attack \predator
        \end{equation}
        Similarly, for the hydrogen fleet:

        \begin{equation}
            \gpredator \leftarrow \efficiency \prey - \gpredator
        \end{equation}
        Consequently, we propose that the number of cars that can be supported by the infrastructures in conventional fuel and hydrogen can be described by the following model:
        
        \begin{subequations} \label{eq-modif}
  \begin{empheq}[left=\empheqlbrace]{align}
                    \dot{\prey} &= \prey(-\gprey - \attack \predator) +\mu_c \label{eq-lvm_prey}\\
                    \dot{\predator} &= \predator\big(\efficiency \prey - \gpredator \big)+\mu_h \label{eq-lvm_predator}
        \end{empheq}
        \end{subequations}
        
        where $\mu_c$ and $\mu_h$ correspond to the resources for conventional and hydrogen vehicles. 
        The attack $\attack$ and efficiency $\efficiency$ rates translate how the policies influence the number of vehicles.
        In particular, when $\attack=\efficiency$, it means that no new cars is introduced due to external factors (i.e. policies), but that a fraction of the vehicles is simply transitioning from one fleet to the other. 
        In the rest of the manuscript, for simplicity, the model from Eqs.~\eqref{eq-modif} will be referred as \LVM (LVM). 
        The architecture of the model is illustrated Fig.~\ref{fig-abstract}.

%%%%%%%%%%%%%%%%%%%%%%%%%%%%%%%%%%%%%%%%%%%%%%%%%%%%%%%%%%%%%%%%%%%%
%%%%%%%%%%%%%%%%%%%%%%%%%% SECTION %%%%%%%%%%%%%%%%%%%%%%%%%%%%%%%%%
\section{Results and discussions of models}\label{sec-res}
        \subsection{Validity of the growth model}\label{ssec-growth_val}

            The models introduced in Sec.~\ref{sec-model} have been implemented.
            They are being solved using a standard Runge-Kutta 4 scheme.
            To validate the model, we aim at representing the UK market.

            \begin{figure}
            \subfloat[]{
            \begin{tikzpicture} 
                \begin{axis}[
                    /pgf/number format/.cd,
                    use comma,
                    1000 sep={},
                    height = .40\linewidth,
                    width = .50\linewidth,
                    legend pos=north west,
                    legend style={fill=white, fill opacity=0.5, draw opacity=1,text opacity=1},
                    xlabel={years},
                    ylabel={fleet (Mveh)},
                    %ytick = {1.e3,1.e4,1.e6},
                    %xtick = {.1,.5,1,5},
                    %xmin=1970,xmax=2020,
                    %ymin=50,ymax=2e6,
                    ]
                    \addplot[mark=none,color=blue,solid] 
                        coordinates{
                            ( 1960  , 0.38 )
                            ( 1961  , 1.02 )
                            ( 1962  , 1.65 )
                            ( 1963  , 2.27 )
                            ( 1964  , 2.89 )
                            ( 1965  , 3.50 )
                            ( 1966  , 4.11 )
                            ( 1967  , 4.71 )
                            ( 1968  , 5.31 )
                            ( 1969  , 5.90 )
                            ( 1970  , 6.48 )
                            ( 1971  , 7.06 )
                            ( 1972  , 7.63 )
                            ( 1973  , 8.19 )
                            ( 1974  , 8.75 )
                            ( 1975  , 9.31 )
                            ( 1976  , 9.86 )
                            ( 1977  , 10.40 )
                            ( 1978  , 10.94 )
                            ( 1979  , 11.47 )
                            ( 1980  , 12.00 )
                            ( 1981  , 12.52 )
                            ( 1982  , 13.04 )
                            ( 1983  , 13.55 )
                            ( 1984  , 14.06 )
                            ( 1985  , 14.56 )
                            ( 1986  , 15.06 )
                            ( 1987  , 15.55 )
                            ( 1988  , 16.03 )
                            ( 1989  , 16.52 )
                            ( 1990  , 16.99 )
                            ( 1991  , 17.47 )
                            ( 1992  , 17.93 )
                            ( 1993  , 18.40 )
                            ( 1994  , 18.86 )
                            ( 1995  , 19.31 )
                            ( 1996  , 19.76 )
                            ( 1997  , 20.21 )
                            ( 1998  , 20.65 )
                            ( 1999  , 21.08 )
                            ( 2000  , 21.51 )
                            ( 2001  , 21.94 )
                            ( 2002  , 22.36 )
                            ( 2003  , 22.78 )
                            ( 2004  , 23.20 )
                            ( 2005  , 23.61 )
                            ( 2006  , 24.02 )
                            ( 2008  , 24.42 )
                            ( 2009  , 24.82 )
                            ( 2010  , 25.21 )
                            ( 2011  , 25.60 )
                            ( 2012  , 25.99 )
                            ( 2013  , 26.37 )
                            ( 2014  , 26.75 )
                            ( 2015  , 27.13 )
                            ( 2016  , 27.50 )
                            ( 2017  , 27.87 )
                            ( 2018  , 28.23 )
                            ( 2019  , 28.59 )
                            ( 2020  , 28.95 )
                      };\addlegendentry{growth model}
                    \addplot[mark=x,color=red,solid,very thick] 
                        coordinates{
                        (1971,8)
                        (1976,10)
                        (1981,10)
                        (1986,13)
                        (1991,16)
                        (1996,18)
                        (2001,20)
                        (2006,25)
                        (2011,27)
                        (2016,29)
                        };\addlegendentry{data from~\cite{Leibling2008}}
                \addplot[mark=none,color=red,solid,domain=1970:2020,samples=50] {0.4921*x - 963.44}; \addlegendentry{linear model}
                \end{axis}
            \end{tikzpicture}
            }
            \subfloat[]{
            \begin{tikzpicture} 
                \begin{axis}[
                    /pgf/number format/.cd,
                    use comma,
                    1000 sep={},
                    height = .40\linewidth,
                    width = .50\linewidth,
                    legend pos=north west,
                    legend style={fill=white, fill opacity=0.5, draw opacity=1,text opacity=1},
                    xlabel={years},
                    ylabel={fleet (Mveh)},
                    %ytick = {1.e3,1.e4,1.e6},
                    %xtick = {.1,.5,1,5},
                    %xmin=1970,xmax=2020,
                    %ymin=50,ymax=2e6,
                    ]
                    \addplot[mark=none,color=blue,solid] 
                        coordinates{
                            ( 1958  , -0.92 )
                            ( 1959  , -0.27 )
                            ( 1960  , 0.38 )
                            ( 1961  , 1.02 )
                            ( 1962  , 1.65 )
                            ( 1963  , 2.27 )
                            ( 1964  , 2.89 )
                            ( 1965  , 3.50 )
                            ( 1966  , 4.11 )
                            ( 1967  , 4.71 )
                            ( 1968  , 5.31 )
                            ( 1969  , 5.90 )
                            ( 1970  , 6.48 )
                            ( 1971  , 7.06 )
                            ( 1972  , 7.63 )
                            ( 1973  , 8.19 )
                            ( 1974  , 8.75 )
                            ( 1975  , 9.31 )
                            ( 1976  , 9.86 )
                            ( 1977  , 10.40 )
                            ( 1978  , 10.94 )
                            ( 1979  , 11.47 )
                            ( 1980  , 12.00 )
                            ( 1981  , 12.52 )
                            ( 1982  , 13.04 )
                            ( 1983  , 13.55 )
                            ( 1984  , 14.06 )
                            ( 1985  , 14.56 )
                            ( 1986  , 15.06 )
                            ( 1987  , 15.55 )
                            ( 1988  , 16.03 )
                            ( 1989  , 16.52 )
                            ( 1990  , 16.99 )
                            ( 1991  , 17.47 )
                            ( 1992  , 17.93 )
                            ( 1993  , 18.40 )
                            ( 1994  , 18.86 )
                            ( 1995  , 19.31 )
                            ( 1996  , 19.76 )
                            ( 1997  , 20.21 )
                            ( 1998  , 20.65 )
                            ( 1999  , 21.08 )
                            ( 2000  , 21.51 )
                            ( 2001  , 21.94 )
                            ( 2002  , 22.36 )
                            ( 2003  , 22.78 )
                            ( 2004  , 23.20 )
                            ( 2005  , 23.61 )
                            ( 2006  , 24.02 )
                            ( 2008  , 24.42 )
                            ( 2009  , 24.82 )
                            ( 2010  , 25.21 )
                            ( 2011  , 25.60 )
                            ( 2012  , 25.99 )
                            ( 2013  , 26.37 )
                            ( 2014  , 26.75 )
                            ( 2015  , 27.13 )
                            ( 2016  , 27.50 )
                            ( 2017  , 27.87 )
                            ( 2018  , 28.23 )
                            ( 2019  , 28.59 )
                            ( 2020  , 28.95 )
                            ( 2021  , 29.30 )
                            ( 2022  , 29.65 )
                            ( 2023  , 30.00 )
                            ( 2024  , 30.34 )
                            ( 2025  , 30.68 )
                            ( 2026  , 31.02 )
                            ( 2027  , 31.35 )
                            ( 2028  , 31.68 )
                            ( 2029  , 32.01 )
                            ( 2030  , 32.33 )
                            ( 2031  , 32.65 )
                            ( 2032  , 32.97 )
                            ( 2033  , 33.28 )
                            ( 2034  , 33.59 )
                            ( 2035  , 33.90 )
                            ( 2036  , 34.20 )
                            ( 2037  , 34.51 )
                            ( 2038  , 34.80 )
                            ( 2039  , 35.10 )
                            ( 2040  , 35.39 )
                            ( 2041  , 35.68 )
                            ( 2042  , 35.97 )
                            ( 2043  , 36.25 )
                            ( 2044  , 36.53 )
                            ( 2045  , 36.81 )
                            ( 2046  , 37.09 )
                            ( 2047  , 37.36 )
                            ( 2048  , 37.63 )
                            ( 2049  , 37.90 )
                            ( 2050  , 38.16 )
                            ( 2051  , 38.42 )
                            ( 2052  , 38.68 )
                            ( 2053  , 38.94 )
                            ( 2054  , 39.19 )
                            ( 2055  , 39.44 )
                            ( 2056  , 39.69 )
                            ( 2057  , 39.94 )
                            ( 2058  , 40.18 )
                            ( 2059  , 40.43 )
                            ( 2060  , 40.67 )
                            ( 2061  , 40.90 )
                            ( 2062  , 41.14 )
                            ( 2063  , 41.37 )
                            ( 2064  , 41.60 )
                            ( 2065  , 41.83 )
                            ( 2066  , 42.05 )
                            ( 2067  , 42.28 )
                            ( 2068  , 42.50 )
                            ( 2069  , 42.72 )
                            ( 2070  , 42.93 )
                            ( 2071  , 43.15 )
                            ( 2072  , 43.36 )
                            ( 2073  , 43.57 )
                            ( 2074  , 43.78 )
                            ( 2075  , 43.98 )
                            ( 2076  , 44.19 )
                            ( 2077  , 44.39 )
                            ( 2078  , 44.59 )
                            ( 2079  , 44.79 )
                            ( 2080  , 44.98 )
                            ( 2081  , 45.18 )
                            ( 2082  , 45.37 )
                            ( 2083  , 45.56 )
                            ( 2084  , 45.75 )
                            ( 2085  , 45.94 )
                            ( 2086  , 46.12 )
                            ( 2087  , 46.30 )
                            ( 2088  , 46.48 )
                            ( 2089  , 46.66 )
                            ( 2090  , 46.84 )
                            ( 2091  , 47.02 )
                            ( 2092  , 47.19 )
                            ( 2093  , 47.36 )
                            ( 2094  , 47.53 )
                            ( 2095  , 47.70 )
                            ( 2096  , 47.87 )
                            ( 2097  , 48.03 )
                            ( 2098  , 48.20 )
                            ( 2099  , 48.36 )
                            ( 2100  , 48.52 )
                            ( 2101  , 48.68 )
                            ( 2102  , 48.84 )
                            ( 2103  , 48.99 )
                            ( 2104  , 49.15 )
                            ( 2105  , 49.30 )
                            ( 2106  , 49.45 )
                            ( 2107  , 49.60 )
                            ( 2108  , 49.75 )
                            ( 2109  , 49.89 )
                            ( 2110  , 50.04 )
                            ( 2111  , 50.18 )
                            ( 2112  , 50.33 )
                            ( 2113  , 50.47 )
                            ( 2114  , 50.61 )
                            ( 2115  , 50.74 )
                            ( 2116  , 50.88 )
                            ( 2117  , 51.02 )
                            ( 2118  , 51.15 )
                            ( 2119  , 51.28 )
                            ( 2120  , 51.41 )
                            ( 2121  , 51.54 )
                            ( 2122  , 51.67 )
                            ( 2123  , 51.80 )
                            ( 2124  , 51.93 )
                            ( 2125  , 52.05 )
                            ( 2126  , 52.18 )
                            ( 2127  , 52.30 )
                            ( 2128  , 52.42 )
                            ( 2129  , 52.54 )
                            ( 2130  , 52.66 )
                            ( 2131  , 52.78 )
                            ( 2132  , 52.89 )
                            ( 2133  , 53.01 )
                            ( 2134  , 53.12 )
                            ( 2135  , 53.24 )
                            ( 2136  , 53.35 )
                            ( 2137  , 53.46 )
                            ( 2138  , 53.57 )
                            ( 2139  , 53.68 )
                            ( 2140  , 53.78 )
                            ( 2141  , 53.89 )
                            ( 2142  , 54.00 )
                            ( 2143  , 54.10 )
                            ( 2144  , 54.20 )
                            ( 2145  , 54.31 )
                            ( 2146  , 54.41 )
                            ( 2147  , 54.51 )
                            ( 2148  , 54.61 )
                            ( 2149  , 54.71 )
                            ( 2150  , 54.80 )
                            ( 2151  , 54.90 )
                            ( 2152  , 54.99 )
                            ( 2153  , 55.09 )
                            ( 2154  , 55.18 )
                            ( 2155  , 55.27 )
                            ( 2156  , 55.37 )
                            ( 2157  , 55.46 )
                            ( 2158  , 55.55 )
                            ( 2159  , 55.64 )
                            ( 2160  , 55.72 )
                            ( 2161  , 55.81 )
                            ( 2162  , 55.90 )
                            ( 2163  , 55.98 )
                            ( 2164  , 56.07 )
                            ( 2165  , 56.15 )
                            ( 2166  , 56.23 )
                            ( 2167  , 56.32 )
                            ( 2168  , 56.40 )
                            ( 2169  , 56.48 )
                            ( 2170  , 56.56 )
                            ( 2171  , 56.64 )
                            ( 2172  , 56.72 )
                            ( 2173  , 56.79 )
                            ( 2174  , 56.87 )
                            ( 2175  , 56.94 )
                            ( 2176  , 57.02 )
                            ( 2177  , 57.09 )
                            ( 2178  , 57.17 )
                            ( 2179  , 57.24 )
                            ( 2180  , 57.31 )
                            ( 2181  , 57.38 )
                            ( 2182  , 57.45 )
                            ( 2183  , 57.52 )
                            ( 2184  , 57.59 )
                            ( 2185  , 57.66 )
                            ( 2186  , 57.73 )
                            ( 2187  , 57.80 )
                            ( 2188  , 57.86 )
                            ( 2189  , 57.93 )
                            ( 2190  , 58.00 )
                            ( 2191  , 58.06 )
                            ( 2192  , 58.12 )
                            ( 2193  , 58.19 )
                            ( 2194  , 58.25 )
                            ( 2195  , 58.31 )
                            ( 2196  , 58.37 )
                            ( 2197  , 58.43 )
                            ( 2198  , 58.49 )
                            ( 2199  , 58.55 )
                            ( 2200  , 58.61 )
                            ( 2201  , 58.67 )
                            ( 2202  , 58.73 )
                            ( 2203  , 58.79 )
                            ( 2204  , 58.84 )
                            ( 2205  , 58.90 )
                            ( 2206  , 58.96 )
                            ( 2207  , 59.01 )
                            ( 2208  , 59.07 )
                            ( 2209  , 59.12 )
                            ( 2210  , 59.17 )
                            ( 2211  , 59.23 )
                            ( 2212  , 59.28 )
                            ( 2213  , 59.33 )
                            ( 2214  , 59.38 )
                            ( 2215  , 59.43 )
                            ( 2216  , 59.48 )
                            ( 2217  , 59.53 )
                            ( 2218  , 59.58 )
                            ( 2219  , 59.63 )
                            ( 2220  , 59.68 )
                            ( 2221  , 59.73 )
                            ( 2222  , 59.77 )
                            ( 2223  , 59.82 )
                            ( 2224  , 59.87 )
                            ( 2225  , 59.91 )
                            ( 2226  , 59.96 )
                            ( 2227  , 60.00 )
                            ( 2228  , 60.05 )
                            ( 2229  , 60.09 )
                            ( 2230  , 60.14 )
                            ( 2231  , 60.18 )
                            ( 2232  , 60.22 )
                            ( 2233  , 60.26 )
                            ( 2234  , 60.31 )
                            ( 2235  , 60.35 )
                            ( 2236  , 60.39 )
                            ( 2237  , 60.43 )
                            ( 2238  , 60.47 )
                            ( 2239  , 60.51 )
                            ( 2240  , 60.55 )
                            ( 2241  , 60.59 )
                            ( 2242  , 60.63 )
                            ( 2243  , 60.67 )
                            ( 2244  , 60.70 )
                            ( 2245  , 60.74 )
                            ( 2246  , 60.78 )
                            ( 2247  , 60.82 )
                            ( 2248  , 60.85 )
                            ( 2249  , 60.89 )
                            ( 2250  , 60.92 )
                            ( 2251  , 60.96 )
                            ( 2252  , 61.00 )
                            ( 2253  , 61.03 )
                            ( 2254  , 61.06 )
                            ( 2255  , 61.10 )
                            ( 2256  , 61.13 )
                            ( 2257  , 61.17 )
                            ( 2258  , 61.20 )
                      };\addlegendentry{growth model}
            %        \addplot[mark=x,color=red,solid,very thick] 
            %            coordinates{
            %            (1971,8)
            %            (1976,10)
            %            (1981,10)
            %            (1986,13)
            %            (1991,16)
            %            (1996,18)
            %            (2001,20)
            %            (2006,25)
            %            (2011,27)
            %            (2016,29)
            %            };
                \addplot[mark=none,color=red,solid,domain=1970:2200,samples=50] {0.4921*x - 963.44}; \addlegendentry{linear model}
                \end{axis}
            \end{tikzpicture}
            }
            \caption{UK fleet size simulated with the growth model. a): Comparison with data. b): long term prediction}
            \label{fig-growth_fit}
            \end{figure}
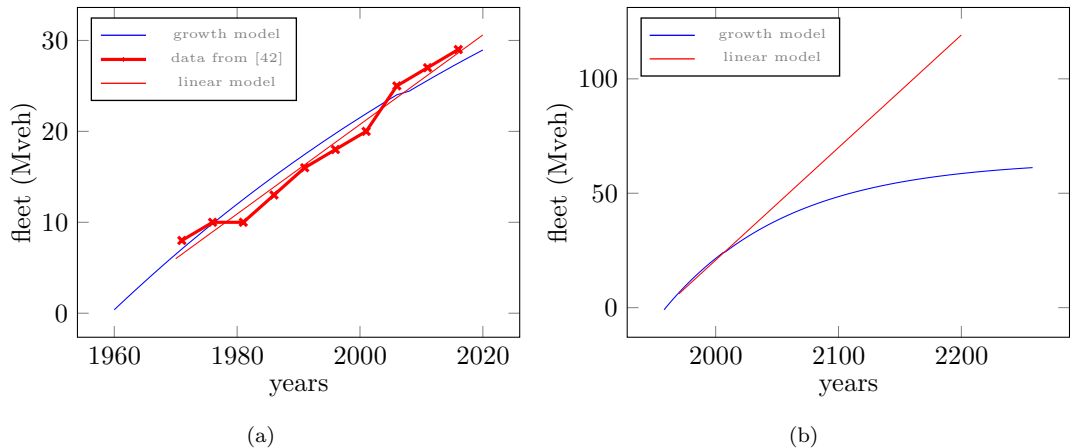

        The first order growth model developed in Sec.~\ref{sec-model} was simulated in Fig.~\ref{fig-growth_fit}. 
        The parameters used are 
        $\gamma = -0.01\,year^{-1}$ 
        and 
        $\mu = 0.65\,Mveh.year^{-1}$. 
        These parameters were identified using a standard fit procedure, against the UK’s road vehicles data from~\cite{Leibling2008}.
        
        %Fig.~\ref{fig-growth} shows that the simulated model projects the growth as demonstrated in Fig.~\ref{fig-data_growth}. 
        
        The point-wise error between a data point $x_{data}$ and a point produced by the model $x_{model}$ is defined as:
        \begin{equation}
            e(x_{data},x_{model}):=\left|\dfrac{x_{data}-x_{model}}{x_{data}}\right|.
        \end{equation}
        %\todo{figure seems different, with a first order dynamics compared to a linear dynamics of $\approx +5M/decade$. To review with Sahdia}

        The growth of conventional vehicles is compared to the growth model, Fig.~\ref{fig-growth_fit}. 
        The average point-wise error is $<e>=0.07\% \pm 0.03\%$.
        There is a strong agreement;
        it indicates that the model developed captures the UK’s growth.

        The model is able to reasonably predict the UK’s growth for conventional vehicles.
        The main feature of the model is that it saturates with a total fleet of $\mu/\gamma = 65Mveh$, and around $48.5Mveh$ in 2100.
        Assuming a population of $81M$ in 2100\footnote{2019 Revision of World Population Prospects, \url{https://population.un.org/wpp/}}, it means around 600 cars per thousand people.
        Not only the model matches the Gompertz saturation curve, but it also reproduces the numbers in~\cite{Dargay2007}, which aligns with a saturation of two cars per household, or around 700 to 800 cars per thousand people,~\cite{Rota2016}.

    %!!!!!!!!!!!!!!!!!!!!!!!!!! SUBSECTION !!!!!!!!!!!!!!!!!!!!!!!!!!
    \subsection{Analytical analysis of the \LVM}
        In this section, we investigate the mathematical properties of the model Eq.~\eqref{eq-modif}.
        
        \subsubsection{Significance of the model parameters}
        The significance of the parameters is fundamental when designing a model.
        
        The growth rate parameters $\gpredator,\gprey$ are related to the settling time. 
        It describes the dynamics of the organic growth of the fleet driven by the available resources $\mu_c,\mu_h$.
        The efficiency and attack rates $\efficiency,\attack$ are mathematically related to the interactions between the fleets. 
        Practically, it translates the incentive of switching from a type of vehicle to another type of vehicle.

    \subsubsection{Mathematical properties}        
        An important properties is the global growth of the model.
        From Eq.~\eqref{eq-modif}:
        \begin{equation}
                \dot{\prey} + \dot{\predator} 
                    =
                     \prey \predator (\efficiency-\attack) -( \gprey \prey  +\gpredator \predator) + \mu_h +\mu_c
            \label{eq-as-model}
        \end{equation}
        For developing the intuition, let's temporary assume that $\attack=\efficiency$ and $\gpredator=\gprey = \gamma$.
        Noting $N=\prey+\predator$ and $\mu = \mu_h + \mu_c$, then Eq.~\eqref{eq-as-model} becomes:
        $$ 
        \dot{N} = \mu - \gamma N
        $$
        The LVM collapses to the first order model growth for the fleet of vehicle (see Eq.~\eqref{eq-grate}), and the total number of vehicles should be, as in Eq.~\eqref{eq-limit}:
        
        \begin{equation}
            N_{\infty} = \dfrac{\mu}{\gamma}
            \label{eq-limit_total}
        \end{equation}
        Interestingly, the final fleet total is independant from the attack and efficiency rates.

        It means the proposed \LVM captures the following properties:
        \begin{itemize}
            \item the growth then saturation of the total fleet of vehicle;
            \item the competition between the type of vehicles;
            \item the dynamics of the fleet associated to each type of vehicle. A growth then saturation of the fleet, a transitional growth then decrease of the fleet, or simply a decline of the fleet are possible, depending on the set of parameters.
        \end{itemize}

        In the same spirit as Eq.~\eqref{eq-limit}, the model Eq.~\ref{eq-modif} reaches asymptotic values for $\prey$ and $\predator$.
        Putting the derivatives to zero and rearranging lead to:
        \begin{equation}
            {\prey}_{\infty} =\dfrac{a\mu_h + \epsilon \mu_c + \gamma_c\gamma_h - \sqrt{a^2 \mu_h^2 + 2 a \epsilon \mu_c \mu_h + 2 a \gamma_c \gamma_h \mu_h + \epsilon^2 \mu_c^2 - 2\epsilon\gamma_c\gamma_h\mu_c + \gamma_c^2 \gamma_h^2}}{2 \epsilon\gamma_c}
            \label{eq-xinf}
        \end{equation}
        \begin{equation}
            {\predator}_{\infty} =\dfrac{a \mu_h + \epsilon \mu_c - \gamma_c \gamma_h + \sqrt{a^2 \mu_h^2 + 2 a \epsilon \mu_c \mu_h + 2 a \gamma_c \gamma_h \mu_h + \epsilon^2 \mu_c^2 - 2 \epsilon \gamma_c \gamma_h \mu_c + \gamma_c^2 \gamma_h^2}}{2 a \gamma_h}
            \label{eq-yinf}
        \end{equation}
        
        The asymptotic solution reveals important interactions between the prey and the predator. 
        %It is interesting to see that, o
        Similarly to the first order model, the quantities $\mu_h/\gpredator$ and $\mu_c/\gprey$ play an role in the asymptotic value.
        However, there is several other terms that can change the equilibrium point.
        The behavior of the asymptotic solution is related to the sign of the argument $\Delta$ of the square root:
        \begin{equation}
        \Delta:=a^2 \mu_h^2 + 2 a \epsilon \mu_c \mu_h + 2 a \gamma_c \gamma_h \mu_h + \epsilon^2 \mu_c^2  + \gamma_c^2 \gamma_h^2 - 2 \epsilon \gamma_c \gamma_h \mu_c
            \label{eq-delta}
        \end{equation}
        If $\Delta$ is positive, then the asymptotic solution is a constant.
        If it is negative, then the solution will oscillate, as in Fig.~\ref{fig-pp}.

        Rearranging positive and negative terms of the argument leads to:
        \begin{equation}
            a^2 \mu_h^2 + 2 a \epsilon \mu_c \mu_h + 2 a \gamma_c \gamma_h \mu_h + \epsilon^2 \mu_c^2  + \gamma_c^2 \gamma_h^2 > 2 \epsilon \gamma_c \gamma_h \mu_c
            \label{eq-stab}
        \end{equation}

        The sign, in Eq.~\eqref{eq-stab}, is determined by the right hand side: the efficiency $\efficiency$, the growth rates, and the support received by the conventional vehicle chain supply $\mu_c$.
        Intuitively, as long as there is enough drive for conventional vehicles, there will be burst of growth for this fleet, and the entire system will be destabilized.

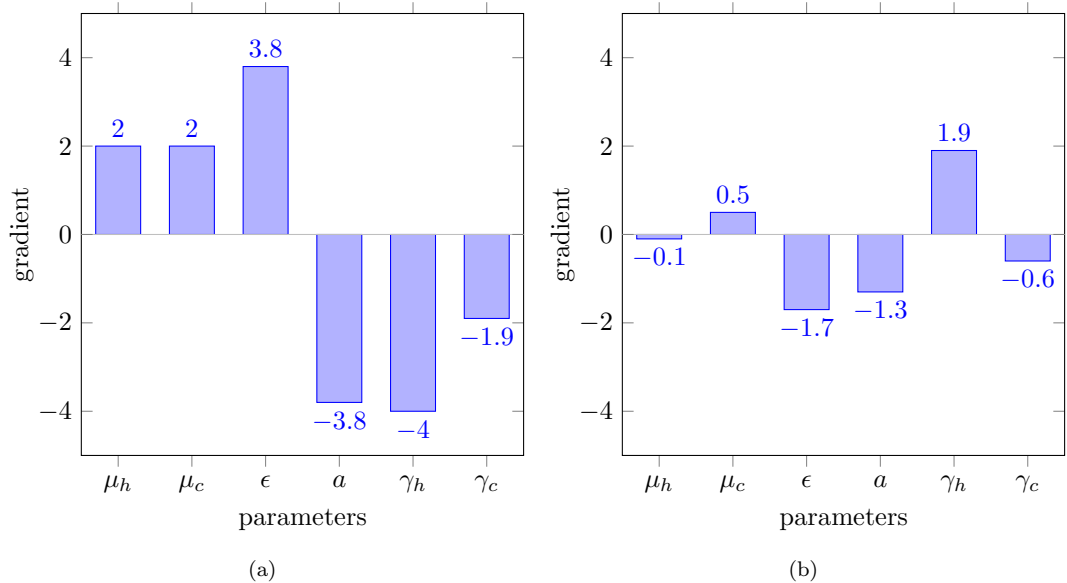
\begin{figure}\center
            \pgfplotsset{xticklabel={\tick},scaled x ticks=false}
            \pgfplotsset{plot coordinates/math parser=false}\subfloat[]{
            \begin{tikzpicture}
                \begin{axis}[
                    extra y ticks = 0,
                    extra y tick labels = ,
                    extra y tick style = {grid = major},
                    nodes near coords,
                    nodes near coords align={vertical},
                    ybar,bar width=17pt,
                    height=0.5\textwidth,
                    width=0.5\textwidth,
                    symbolic x coords={$\mu_h$,$\mu_c$,$\epsilon$,$a$,$\gamma_h$,$\gamma_c$},
                    xlabel={parameters},
                    ylabel={gradient},
                    xtick=data,
                    ymin=-5,
                    ymax=5,
                    ]
                    \addplot table[x=param, y=gradh] {./data/gradients2.dat};
                \end{axis}
            \end{tikzpicture}
            \label{sfig-gradh}
            }
            \subfloat[]{
                \begin{tikzpicture}
                \begin{axis}[
                    extra y ticks = 0,
                    extra y tick labels = ,
                    extra y tick style = {grid = major},
                    nodes near coords,
                    nodes near coords align={vertical},
                    ybar,bar width=17pt,
                    height=0.5\textwidth,
                    width=0.5\textwidth,
                    symbolic x coords={$\mu_h$,$\mu_c$,$\epsilon$,$a$,$\gamma_h$,$\gamma_c$},
                    xlabel={parameters},
                    ylabel={gradient},
                    xtick=data,
                    ymin=-5,
                    ymax=5,
                    ]
                    \addplot table[x=param, y=gradc] {./data/gradients2.dat};
                \end{axis}
            \end{tikzpicture}
            \label{sfig-gradc}
            }
            \caption{Gradients of the asymptotic values (signed pseudo log10). (a): for the hydrogen fleet. (b): for the conventional fleet. Value used can be found Tab.~\ref{tab-grad}}
            \label{fig-gradients}
        \end{figure}

        The sensibility can also be studied, by calculating the gradient of the asymptotic solutions with respect to the parameters. 
        The sensibility allows to predict how varying parameters affect the outcome of the system.
        The sensitivity has been plotted Fig.~\ref{fig-gradients}.
        Practically, it indicates how policies can influence the fleet numbers. 
        %The two parameters that can be impacted are mostly the supply chains parameters $\mu_c$ and $\mu_h$.

        Differentiation of the asymptotic value leads to Eqs.~\eqref{eq-grad_hydro} for the hydrogen fleet, and to Eqs.~\eqref{eq-grad_conv} for the conventional fleet. 
        The notation $\Delta$ has been kept for the sake of readability.

        \begin{subequations}\label{eq-grad_hydro}
  \begin{empheq}[]{align}
              \dfrac{\partial {\predator}_{\infty}}{\partial \mu_h} &= 
        \dfrac{a \mu_h + \epsilon \mu_c + \gamma_c \gamma_h + \sqrt{\Delta}}{2 \gamma_h \sqrt{\Delta}}
        \label{eq-dxdmuh} \\
          \dfrac{\partial {\predator}_{\infty}}{\partial \mu_c} &= 
        \epsilon \dfrac{a \mu_h + \epsilon \mu_c - \gamma_c \gamma_h + \sqrt{\Delta}}{2 a \gamma_h \sqrt{\Delta}}
        \label{eq-dxdmuc} \\
          \dfrac{\partial {\predator}_{\infty}}{\partial \epsilon} &=     
                \mu_c \dfrac{a \mu_h + \epsilon \mu_c - \gamma_c \gamma_h + \sqrt{\Delta}}
                {2 a \gamma_h \sqrt{\Delta}}
        \label{eq-dxdeps}\\
          \dfrac{\partial {\predator}_{\infty}}{\partial \attack} &= 
-\dfrac{
        a \epsilon \mu_c \mu_h + a \gamma_c \gamma_h \mu_h + \epsilon^2 \mu_c^2 - 2 \epsilon \gamma_c \gamma_h \mu_c 
        + \gamma_c^2 \gamma_h^2 
        +(\epsilon \mu_c - \gamma_c \gamma_h) \sqrt{\Delta}}
{2 a^2 \gamma_h \sqrt{\Delta}
}
        \label{eq-dxda}\\
          \dfrac{\partial {\predator}_{\infty}}{\partial \gpredator} &= 
-\dfrac{
    a^2 \mu_h^2 + 2 a \epsilon \mu_c \mu_h + a \gamma_c \gamma_h \mu_h 
    + \epsilon^2 \mu_c^2 - \epsilon \gamma_c \gamma_h \mu_c 
    + (a \mu_h + \epsilon \mu_c ) \sqrt{\Delta}
}{
2 a \gamma_h^2 \sqrt{\Delta}
}
        \label{eq-dxdgammah}\\
          \dfrac{\partial {\predator}_{\infty}}{\partial \gprey} &=
\dfrac{(a \mu_h - \epsilon \mu_c + \gamma_c \gamma_h - \sqrt{\Delta})
}{
2 a \sqrt{\Delta}
}
        \label{eq-dxdgammac}
        \end{empheq}
        \end{subequations}

        \begin{subequations}\label{eq-grad_conv}
  \begin{empheq}[]{align}
          \dfrac{\partial {\prey}_{\infty}}{\partial \mu_h} &= 
        a \dfrac{-a \mu_h - \epsilon \mu_c - \gamma_c \gamma_h + \sqrt{\Delta}}{2 \epsilon \gamma_c \sqrt{\Delta}}
        \label{eq-dydmuh} \\
               \dfrac{\partial {\prey}_{\infty}}{\partial \mu_c} &= 
            \dfrac{-a \mu_h - \epsilon \mu_c + \gamma_c \gamma_h + \sqrt{\Delta}}{2 \gamma_c \sqrt{\Delta}}
        \label{eq-dydmuc}\\
            \dfrac{\partial {\prey}_{\infty}}{\partial \epsilon} &= 
\dfrac{
        a^2 \mu_h^2 + a \epsilon \mu_c \mu_h + 2 a \gamma_c \gamma_h \mu_h
        - \epsilon \gamma_c \gamma_h \mu_c + \gamma_c^2 \gamma_h^2 
        -( a \mu_h + \gamma_c \gamma_h )  \sqrt{\Delta}
}{
2 \epsilon^2 \gamma_c \sqrt{\Delta}}
   \label{eq-dydeps}\\
      \dfrac{\partial {\prey}_{\infty}}{\partial \attack} &=
       \mu_h 
       \dfrac{-a \mu_h - \epsilon \mu_c - \gamma_c \gamma_h + \sqrt{\Delta}
       }
       {2 \epsilon \gamma_c \sqrt{\Delta}
       }
   \label{eq-dyda}\\
      \dfrac{\partial {\prey}_{\infty}}{\partial \gpredator} &=
\dfrac{
-a \mu_h + \epsilon \mu_c - \gamma_c \gamma_h + \sqrt{\Delta}
}
{
2 \epsilon \sqrt{\Delta}
}
   \label{eq-dydgammah} \\
    \dfrac{\partial {\prey}_{\infty}}{\partial \gprey} &=
 \dfrac{
 a^2 \mu_h^2 + 2 a \epsilon \mu_c \mu_h + a \gamma_c \gamma_h \mu_h 
 + \epsilon^2 \mu_c^2 - \epsilon \gamma_c \gamma_h \mu_c 
 - (a \mu_h + \epsilon \mu_c) \sqrt{\Delta}
}
 {
 2 \epsilon \gamma_c^2 \sqrt{\Delta}
 }
   \label{eq-dydgammac}
        \end{empheq}
        \end{subequations}

        Looking at the terms $a \mu_h$ and $\efficiency \mu_c$ develop the intuition concerning the effect of the availability of resources (i.e., the supply chains) on the fleet.
        In Eqs.~\eqref{eq-dxdmuh} and~\eqref{eq-dxdmuc}, both terms are positive. 
        It means the supply chains drive the augmentation in the final number of hydrogen vehicles. 
        On the other hand, in Eqs.~\eqref{eq-dydmuh} and~\eqref{eq-dydmuc}, both terms are negative. 
        It means the chain supplies tend to impact negatively the number of conventional vehicle. 
        It is reflected in the first two bars in Fig.~\ref{fig-gradients}.
        It is counter-intuitive to see that the chain supply for conventional vehicle feeds more the growth of the hydrogen vehicle. 
        But it is intrinsic to a predator-prey model: more preys mean a favorable ground for the predators.
        In general, more available resources means more desires from the consumer, and, ultimately, more chances for the long-term dominant competitor to enforce its position.
        
        One final intuition can be developed concerning the importance of the attack rate $\attack$. 
        A larger attack rate seems detrimental to the final value of the hydrogen fleet. 
        It is related to the sign of $\attack$ in Eq.~\eqref{eq-grad_conv}. 
        A larger attack rate reduces the population of vehicles that hydrogen vehicles can predate on.

        \begin{table}\center
            \caption{Parameters used for calculating the gradient of Fig.~\ref{fig-gradients} }\label{tab-grad}
            \begin{tabular}{l|cccccc}
            parameter & $\mu_h$ & $\mu_c$ & $\efficiency$ & $\attack$ & $\gamma_h$ & $\gamma_c$ \\
             & \small $(Mveh/year)$ & \small $(Mveh/year)$ & \small $(Mveh^{-1}year^{-1})$ & \small $(Mveh^{-1}year^{-1})$ & \small $(year^{-1})$ & \small $(year^{-1})$ \\
            \hline
            value & 0.65 & 0.65 & 0.01 & 0.01 & 0.01 & 0.01 
            \end{tabular}
        \end{table}

    %!!!!!!!!!!!!!!!!!!!!!!!!!! SUBSECTION !!!!!!!!!!!!!!!!!!!!!!!!!!        
    \section{Case Study: UK market and policies}\label{sec-policies}
        This section compares the simulation results in terms of vehicle fleet composition for different transition scenarios. 
        A cost analysis of the hydrogen supply chain, based on the most likely scenario, follows. 
        This section concludes on policy recommendations.
        
        \subsection{Scenarios and Dynamics of fleet evolution}\label{ssec-scenarios}
        %\change{In the section 3.3. it starts Three scenario have been designed: Try to provide some introduction as it starts really out of the blue with scenarios. I think it would be better to introduce the policies fist and then say the scenarios. need a little more elaboration/clarification} 

         In 2017, the UK government announced that it will end the sale of all new conventional petrol and diesel cars and vans by 2040,~\cite{act2017}.
         The calendar has recently been accelerated to 2030.
         Further announcements by the government stated that all new vehicles and vans by 2035 will be zero emissions,~\cite{Roadtozero}.

        Three scenario have been designed, corresponding to policies for transitioning to hydrogen transport system.
        Parameters are found Tab.~\ref{tab-values}. 
        The growth rates are kept at $\gpredator=\gprey=0.01year^{-1}$, following the global results of Fig.~\ref{fig-growth_fit}. 
        It means saturation of the fleet takes around 100 years. 
        It will not be shown as the final values are described in Eqs.~\eqref{eq-grad_hydro} and~\eqref{eq-grad_conv}.
        To facilitate the analysis, it is assumed that $\attack=\efficiency$. 
        Practically, it means that the overall growth of the fleet is only driven by the resources $\mu_c$ and $\mu_h$ (see Eq.~\eqref{eq-as-model}), and that individuals are incentivized to change cars, not to have more extra cars.
        The values of $\attack$ and $\efficiency$ reflect the level of incentivization to change vehicles.
        The first scenario (named Low) translates a business-as-usual policy. The available resources are kept low, and the policies for changing vehicles are not incentivized.
        For the second scenario (named Moderate), the parameters have been fitted in order to match the road-to-zero policy,~\cite{Roadtozero}.
        The third scenario (named Aggressive) consider reaching the road-to-zero objectives by 2030.

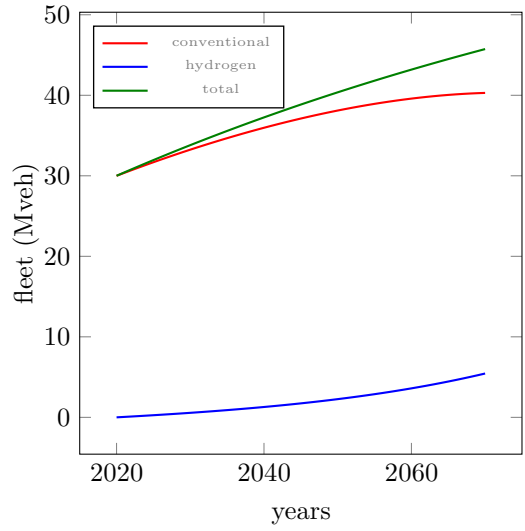
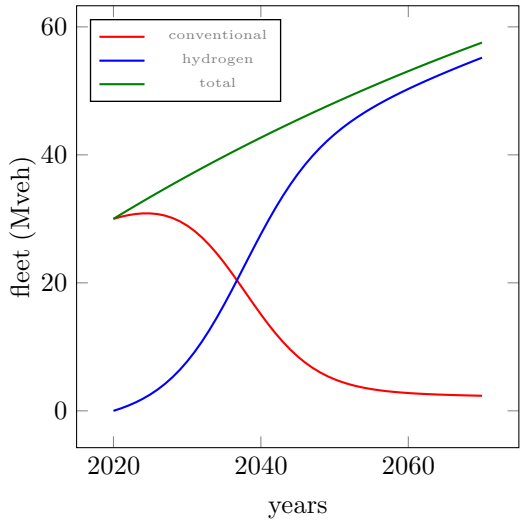
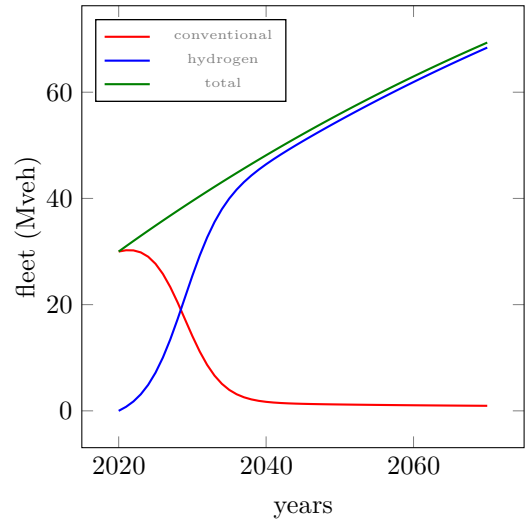
\begin{figure}\center
        \subfloat[]{
                \begin{tikzpicture}
                \begin{axis}[
                    /pgf/number format/.cd,
                    use comma,
                    1000 sep={},
                    height=0.5\textwidth,
                    width=0.5\textwidth,
                    table/col sep=comma,
                    legend pos=north west,
                    legend style={fill=white, fill opacity=0.5, draw opacity=1,text opacity=1},
                    %coords={$\mu_h$,$\mu_c$,$\epsilon$,$a$,$\gamma_h$,$\gamma_c$},
                    %enlarge x limits=0.2,
                    xlabel={years},
                    ylabel={fleet (Mveh)},
                    ]
                    \addplot[thick,mark=none,color=red,solid] table[x=time, y=conv] {./data/hydro_article_scenario_low.csv}; \addlegendentry{conventional}
                    \addplot[thick,mark=none,color=red,blue] table[x=time, y=hydro] {./data/hydro_article_scenario_low.csv}; \addlegendentry{hydrogen}
                    \addplot[thick,mark=none,color=black!50!green,solid] table[x=time, y=total] {./data/hydro_article_scenario_low.csv}; \addlegendentry{total}
                \end{axis}
            \end{tikzpicture}\label{sfig-scenario1}
}

 \subfloat[]{           
                \begin{tikzpicture}
                \begin{axis}[
                    /pgf/number format/.cd,
                    use comma,
                    1000 sep={},
                    height=0.5\textwidth,
                    width=0.5\textwidth,
                    table/col sep=comma,
                    legend pos=north west,
                    legend style={fill=white, fill opacity=0.5, draw opacity=1,text opacity=1},
                    %coords={$\mu_h$,$\mu_c$,$\epsilon$,$a$,$\gamma_h$,$\gamma_c$},
                    %enlarge x limits=0.2,
                    xlabel={years},
                    ylabel={fleet (Mveh)},
                    ]
                    \addplot[thick,mark=none,color=red,solid] table[x=time, y=conv] {./data/hydro_article_scenario_mid.csv}; \addlegendentry{conventional}
                    \addplot[thick,mark=none,color=red,blue] table[x=time, y=hydro] {./data/hydro_article_scenario_mid.csv}; \addlegendentry{hydrogen}
                    \addplot[thick,mark=none,color=black!50!green,solid] table[x=time, y=total] {./data/hydro_article_scenario_mid.csv}; \addlegendentry{total}
                \end{axis}
            \end{tikzpicture}\label{sfig-scenario2}
}
 \subfloat[]{    
                \begin{tikzpicture}
                \begin{axis}[
                    /pgf/number format/.cd,
                    use comma,
                    1000 sep={},
                    height=0.5\textwidth,
                    width=0.5\textwidth,
                    table/col sep=comma,
                    legend pos=north west,
                    legend style={fill=white, fill opacity=0.5, draw opacity=1,text opacity=1},
                    %coords={$\mu_h$,$\mu_c$,$\epsilon$,$a$,$\gamma_h$,$\gamma_c$},
                    %enlarge x limits=0.2,
                    xlabel={years},
                    ylabel={fleet (Mveh)},
                    ]
                    \addplot[thick,mark=none,color=red,solid] table[x=time, y=conv] {./data/hydro_article_scenario_high.csv}; \addlegendentry{conventional}
                    \addplot[thick,mark=none,color=red,blue] table[x=time, y=hydro] {./data/hydro_article_scenario_high.csv}; \addlegendentry{hydrogen}
                    \addplot[thick,mark=none,color=black!50!green,solid] table[x=time, y=total] {./data/hydro_article_scenario_high.csv}; \addlegendentry{total}
                \end{axis}
            \end{tikzpicture}\label{sfig-scenario3}
}
            \caption{Evolution of fleets. a): Low scenario. b): Moderate scenario. c): Aggressive scenario. Values can be found Tab.~\ref{tab-values}}
            \label{fig-scenarios}
        \end{figure}

        Simulations results are plotted Fig.~\ref{fig-scenarios}.
        Total growth of fleets are compatible with the expected trends, as seen Figs.~\ref{fig-growth_fit}, and both the Low and Moderate scenarios produce values similar to the ones reported in~\cite{Brand2020}, as illustrated in Tab.~\ref{tab-verif}

        \begin{table}\center
            \caption{Comparison of scenario outputs for the share of \ZEVs in the fleet composition in 2050}\label{tab-verif}
            \begin{tabular}{l|cccccc}
            Scenario & present study & literature value & Refs \\
            \hline
            Low Scenario & 10\% & 10\% & ~\cite{Brand2020}\\            
            Moderate Scenario & 92\% & 100\% &~\cite{act2019}\\            
            \end{tabular}
        \end{table}

        As expected, if the hydrogen resources are low and if the incentives are low, the growth of the hydrogen fleet is slow, see Fig.~\ref{sfig-scenario1}.
        Most of the fleet remain conventional.
        The third scenario (Fig.~\ref{sfig-scenario3}), with an aggressive policies toward transionning from ICEVs to HEVs (translated by a high attach $\attack$ and efficency $\efficiency$ rates) exhibits a sharp decline in the conventional fleet, with barely conventional vehicles by 2035.
        %The moderate scenario seems the most realistic and can be found Fig.~\ref{sfig-scenario2}.
        The available resources in hydrogen does not yet match the conventional fuel ones, and it will take probably decades, even with strong policies, to fully transition.
        It is encompassed in the Moderate scenario, with $\mu_h \approx \dfrac{1}{2} \mu_c$. 
        It takes around 20 years for the HEV fleet to reach initial ICEV fleet, and the conventional fleet seems to finally disappear only around 2050.

        \subsection{Consequences on the hydrogen supply chain}
        Identifying valid parameters based on the model allows to design and prepare the policies and investments. 
        For instance, accordingly to the Moderate scenario, it means that around $0.35Mveh$ new hydrogen vehicle have to be absorbed each year in the UK market.
        
        \begin{table}\center
            \caption{Values of parameters used in this study}\label{tab-values}
            \begin{tabular}{rl||ccc}
                Parameter &  Definition & Low & Moderate & Aggressive \\
                \hline
                $\gamma_c$ & Growth rate of ICEVs ($year^{-1}$)&
                    $0.01$ & $0.01$ & $0.01$ \\
                $\gamma_h$ & Growth rate of HBVs ($year^{-1}$) &
                    $0.01$ & $0.01$ & $0.01$ \\
                $\attack$ &Attack rate  ($Mveh^{-1}year^{-1}$)&
                $0.001$ & $0.005$ & $0.01$ \\
                $\efficiency$ & Hydrogen efficiency ($Mveh^{-1}year^{-1}$)&
                $0.001$ & $0.005$ & $0.01$ \\
                $\mu_c$ & available resources (fuel) ($Mveh/year$) & 0.65 & 0.65& 0.65\\
                $\mu_h$ & available resources (hydrogen) ($Mveh/year$) & 0.05 & 0.35& 0.65\\
            \end{tabular}
        
        \end{table}

%    \todo{Try to calculate SMR based on scenarios}

        The model, simulated in Sec.~\ref{sec-res}, allows to predict the impact of introducing HBVs as a competitor to conventional vehicles. 
        In the different scenarios, introducing \HFCVs into the road transport market on internal combustion engine vehicles means that hydrogen recharge stations (HRS) have to be deployed to generate and meet the demand.
        %
        %The hydrogen demand can be calculated using two kind of recharge stations (RS). 
        Two archetypical types of stations, one small and one large, are considered for the hydrogen production. Details are found Tab.~\ref{tab-rs}.
        The parameters selected for the refuelling stations are given in Tab.~\ref{tab-storage}.
        An assumption is that HRS operate at $100\%$ efficiency.
        The parameters concerning the capacities of vehicles can be found Tab.~\ref{tab-storage}
        It was assumed that the vehicle would be refuelled once a week when calculating the amount of fuel consumed per year per vehicle. 
        Two kind of HEVs are considered, battery powered EVs with a fuel-cell range extender (HFCRE) and \HFCVs (noted as HFC).
        
        \begin{table}\caption{Production capacities of small (hydrogen solar-powered electrolysis) and large (steam methane reforming) refuelling stations}\label{tab-rs}
        \center
        \begin{tabular}{c|ccc}
            Station type & Capacity ($kg/day$) & Capacity ($kg/year$) & Refs. \\
            \hline
            Small & $200$ & $73000$ & \cite{Campinez2018}\\
            Large & $1000$ & $365000$ &  \cite{Campinez2018}\\
        \end{tabular}
        \end{table}

        \begin{table}\caption{Storage capacity parameters selected for the growth model.}\label{tab-storage}
        \center
        \begin{tabular}{c|ccc}
            Vehicle type & Storage capacity ($kg$) & Fuel consumed & Refs. \\
            &  & per vehicle ($kg/year$) & \\
            \hline
            Hydrogen Fuel Cell & $5$ & $260$ & \cite{Toyota2016,Hyundai2020}  \\
            HFC Range Extender & $1.5$ & $78$ & \cite{Renault2019} \\
        \end{tabular}
        \end{table}

        The moderate scenario from Sec.~\ref{ssec-scenarios} is considered. Around $0.35Mveh$ hydrogen vehicle have to be absorbed each year by the hydrogen supply chain.
        Four transition scenarios are investigated in the following:
                \begin{itemize}
        \item \textbf{S1: Fleet based on HFC, supply chain based on small RS}\\
            A small station can power up to 40 HFC vehicle per year.
            It means that around 8750 small HRS have to be deployed each year, for a total of 262500 deployed by 2050.
        \item \textbf{S2: Fleet based on HFCRE, supply chain based on small RS}\\
            A small station can power up to 133 HFCRE vehicle per year.
            It means that around 2625 small HRS have to be deployed each year, for a total of 78750 deployed by 2050.
         \item \textbf{S3: Fleet based on HFC, supply chain based on large RS}\\
            A large station can power up to 400 HFC vehicle per year.
            It means that around 1750 large HRS have to be deployed each year, for a total of 20010 deployed by 2050.
        \item \textbf{S4: Fleet based on HFCRE, supply chain based on large RS}\\
            A large station can power up to 667 HFCRE vehicle per year.
            It means that around 525 large HRS have to be deployed each year, for a total of 15750 deployed by 2050.
        \end{itemize}
        
        A typical filling station delivers around 5Mkg/year of fuel        \footnote{\url{https://www.racfoundation.org/assets/rac_foundation/content/downloadables/racf_deloitte-fuel_retail-jan13.pdf}}.
        It is 14 times more fuel than the corresponding hydrogen produced by a large HRS. 
        Consequently, the amount of HRS that has to be deployed is equivalent to 1429 fuel filling stations.
        This number can be positively compared to the actual 8000 filling stations in the UK.
        Discrepancies can be explained mostly by the assumption that RS operate at 100\% capacity.
        Filling station are used at high capacity mostly during rush hours, roughly 25\% of the time. 
        At full capacity, around 2000 filling stations would be needed, a much closer figure to the 1429 filling station equivalent.
        Additionally, the efficiency of electric motor is higher than the efficiency of IC engines.
        It validates furthermore the proposed model and scenario.

        %Hydrogen is considered widely to be a clean fuel. 

        \subsection{Policy recommendations for UK's transition to \ZEVs}
        
        \subsection{Preliminary comments}
        The modified LVM proposed allows different penetration strategies and scenarios to be considered over time from a restrictive role to a holistic viewpoint. This is important when determining the role that hydrogen will play. However, the future private vehicle fleet will consist of more vehicle types, rather than simply HBVs and conventional ones. For further research, the two-state model can be extended to encompass other vehicles types to assess the interactions between multiple vehicles types, to outline the role that each will play. This will help assess whether hydrogen will assume a significant role as a HFCVs or constitute a role in the range-extender market. The role of HFCREs will also be interesting seeing as it demands a less exhaustive infrastructure, but a greater challenge for vehicle manufacturers.

        It is to note that hydrogen production may well result in environmental consequences. 
        As the number of plants being built increases, the corresponding environmental impact will also increase.
        Stress on the grid is likely to increase dramatically as well.
        Notably, SMR-based stations will have greater consequences for the environment than a solar-based electrolysis station. 
        It is important that balance is achieved by determining the proportion of each type of station and the overall proportion of role hydrogen will play in the future’s road transportation network.
                %\mescom{With the exception of operating a SMR plant emits a very small amount of emissions.. Meaning?}
        %\mescom{The  emitted is accountable for $99\%$ of the total air emissions and $89\%$ of the system GWP. }
        %Utilising natural gas also results in the emission of methane, which is the second contributor to GWP. 
        %Balancing the energy of the system indicates that $1MJ$ of fossil energy must be consumed for every $0.66MJ$ of hydrogen produced. \mescom{sources? why talking about it ?}
        From an environmental and economic perspective, it is critical to increase the energy efficiencies and ratios of any process leading to a reduction in sources consumed, emissions, wastes and energy consumption,~\cite{Spath2000}.
        %\improvement{ Discuss also some further contribution e.g. policy, and future research opportunities Moreover, it would be nice more discussion of the findings (bring other studies, compare them)} 

        \subsubsection{Cost of the transition} 
        The results of the model indicate, as expected, that two factors are necessary for achieving the road-to-zero objectives.
        
        The first one is the need for a growing hydrogen supply chain.
        The cost to build a small HRS is around \pounds 1million, and the cost for a larger HRS is around  \pounds 5millions,~\cite{Campinez2018}.
        The price per kg of fuel is relatively comparable.
        It means that, at short-term, it is more likely that small-scale refuelling stations will be built, probably directly on filling station sites. 
        For the long-term scenario, larger HRS might be deployed to provide the hydrogen fuel, but it is not expected before a significant part of the fleet composition is based on \HBVs.
        Overall associated infrastructure costs would be around \pounds 2.5billions per year in the most conservative scenario.
        Such changes can only be driven by policy making, as the investments are between \pounds 2.6billion (S2\&S4) and \pounds 8.7billions (S1).
        These significant investments, if carried on by filling stations, will rely on businesses with usually thin margins. 
        Strong public policies, including subsidies, will have to be dedicated to this transition,~\cite{Nocera2016}.
        An integrated approach as part of a national strategy, is needed.
        Public bodies need to manage the general transition, while letting private investors and energy providers leading and deciding local investments.
        
        The second main factor is the exchange rate between ICEVs and HBVs.
        Strong incentive are needed for the population to transition from ICEVs to \ZEVs.
        Costs of alternative fuels vehicles are significantly higher than the costs for conventional ones.
        For allowing these vehicles to penetrate the mass market - a crucial assumption behind the proposed model - public policies are need to push both the research, development of car manufacturers and pushing for an attractive retail price of \HBVs,~\cite{Engstrom2019}.
        To minimize the cost of the infrastructure, it also means that policies should lean toward the development of a HFCRE fleet, rather than a HFC fleet.

        Different studies in the literature have considered the hydrogen infrastructure deployment issue in the UK by assessing either one part of the HSC or a snapshot of the entire supply chain. This paper used a modified LVM to provide a holistic view of the UK’s private vehicle fleet by comparing three penetration strategies for both HFCVs and HFCREs. To capture the transition to a hydrogen-based infrastructure, three demand scenarios were considered derived from both current policies based on decarbonising the private vehicle fleet, and hydrogen-based projects being implemented. The annual hydrogen demand was calculated using the HFCV stock. 

    Current planning by the government has focused on implementing a pre-commercialised infrastructure before expanding on this across important driving routes across the country\footnote{\url{http://www.ukh2mobility.co.uk/the-project/production-and-distribution/}}. As a result, initial smaller HRS can potentially face closure when the need for larger more centralised ones arises. Other studies have indicated that that initially a decentralised infrastructure will be built, and once the uptake of hydrogen has met 15-30\% of the market share, then the strategy of a centralised infrastructure becomes more important,~\cite{Seo2020}. In contrast to the UK, the French government has aligned the development of the hydrogen infrastructure to the demand from different ‘niche applications’ to ensure growth of the infrastructure is matched by the growth of \HFCVs,~\cite{France2018}. 
    One major component of designing a new infrastructure is to ascertain the demand and plan accordingly to avoid under-estimation and overzealous estimations, both resulting in excess costs and consumption of resources. Infrastructure planning should also encompass economies of scale from the outset, outlining a more robust and fit for purpose infrastructure ensuring that \HFCVs plays an important role and not simply a ‘niche-market’ role. Other studies have also cited the importance of economies of scale allowing hydrogen to penetrate the private vehicle market more successfully,~\cite{Tlili2020}. Economies of cost will play a huge role, both, in reducing costs associated with the infrastructure and pushing HFCVs towards a front runner to replace conventional vehicles.

        %The hydrogen capacity of pure hydrogen cars was taken to be 5kg per refill and 1.5kg capacity for a range extender (RE). 
        %The capacity of re 
        %large hydrogen refuelling station is 200kg/day [95]. A HFCV has an onboard storage capacity of 5kg [95,99]. 
        %A station that produces 200kg/day can fuel up to 40 vehicles/day. A 5kg full tank can achieve a range of 200 miles and assume drivers will refuel once a week; an HFCV will require 260kg of fuel/year and an HFC-RE will require 78kg of fuel/year. 

\section{Conclusion} \label{sec-outro}  

    In this paper, a dynamic model based on the \LVM and growth model was formulated and used to simulate the hydrogen demand based on literature. 
    A first order model was developed, and maps the growth of conventional vehicles.
    The growth model developed covers a period of $100$ years and the number of vehicles is similar to those of RAC Foundation from 1971 to current projections. 
    This suggests that the first order model developed can be used to represent the growth of conventional vehicles fleet. 
    Since, the growth of the passenger vehicle fleet in the UK is expected to follow the same trajectory, the model can further be used to predict the number of alternative vehicles replacing the development of conventional ones.
    Then, a two state model was developed. It is able to project what might happen if \HBVs~are introduced.

    The original contributions presented in this work include the application of \LVM to UK’s road transportation. 
    The main results are the prediction of the final composition of the fleet, and a sensitivity analysis on the parameters that influence it. 
    It allows to identify the key parameters that policy makers can use to hasten the transition of the market and the decarbonification of the UK's vehicle fleet.
    In particular, the supply chain in hydrogen is identified as the most important parameter. 
    The cost associated to the deployment of the needed supply chain indicates that strong policies are mandatory for a successful transition.

    %The predator-prey model presented in this work alongside the results are part of the initial work conducted to see if the model is viable and able to validate other results presented in the literature. 
    %This work has demonstrated that the model can be used to represent the UK’s passenger fleet and predict the impact of introducing HBVs. 
    A limitation of the work is that the simulation included assumed HRS operation at $100\%$ efficiency and in reality this is unlikely to be the case. 
    Additionally, the increase in capacity has been assumed constant overtime.
    %However, this was necessary to assess the plausibility of the model against literature, which will then allow different efficiencies and scenarios to be simulated. 
    %The model can be extended to include other variables and predict what will happen using different powertrains. 
    %Another limitation of the current work, and also scope for future work is to integrate and analyse the electrical infrastructure with respect to HFC-RE. 
    %The electrical aspect has not been considered, and so the full picture in terms of fuel economy, and emissions reduced is unclear. 
    %It is unlikely that all or even most of the electricity used to power passenger vehicles will be produced through RES. However, the analysis has highlighted some of the issues and challenges ahead if we are to move to a hydrogen based system. 
    This work is preliminary, and further development of the model and scenarios is being conducted.
    The primary benefit of utilising the \LVM alongside the growth model is a reduction in computing power and complexity in the modelling. 
    Furthermore this reduced the number of parameters in comparison to other models in literature, providing a simpler means of analysing the introduction of hydrogen for passenger vehicles. 
    
    The modified LVM proposed allows different penetration strategies and scenarios to be considered over time from a restrictive role to a holistic viewpoint. This is important when determining the role that hydrogen will play. 
    However, the future private vehicle fleet will consist of more vehicle types, rather than simply \HBVs~and conventional ones. 
    For further research, the two-state model can be extended to encompass other vehicles types to assess the interactions between multiple vehicles types, to outline the role that each will play. 
    This will help assess whether hydrogen will assume a significant role as a \HFCVs~or constitute a role in the range-extender market. 
    The role of HFCREs~will also be interesting seeing as it demands a less exhaustive infrastructure, but a greater challenge for vehicle manufacturers.
    To improve the accuracy of the model, the parameters such as attack rates and growth rates can be empirically measured, and the effect of policies can be measured as well.
    It will allow to i) use time dependent parameters, and ii) identify near optimal policies, for instance using optimal control or model predictive control theories,~\cite{Chu2012}.

\section*{Author contributions}
SM has developped the LVM model and initial methodology and has participated in the writing. FG has developed the formal analysis and has participated in the writing.
\section*{Aknowledgement}
The authors want to thank Dimitra Kalaitzi for fruitful discussions concerning the hydrogen supply chain.
%%
%% End of file `elsarticle-template-3a-num.tex'.
\bibliographystyle{ieeetr}
\bibliography{biblio}

\begin{thebibliography}{10}

\bibitem{IEA2020}
{International Energy Agency}, {\em World energy outlook 2019}.
\newblock Citeseer, 2020.

\bibitem{Haslam2012}
G.~E. Haslam, J.~Jupesta, and G.~Parayil, ``Assessing fuel cell vehicle
  innovation and the role of policy in japan, korea, and china,'' {\em
  International Journal of Hydrogen Energy}, vol.~37, pp.~14612--14623, oct
  2012.

\bibitem{Acar2014}
C.~Acar and I.~Dincer, ``Comparative assessment of hydrogen production methods
  from renewable and non-renewable sources,'' {\em International Journal of
  Hydrogen Energy}, vol.~39, pp.~1--12, jan 2014.

\bibitem{Meckling2019}
J.~Meckling and J.~Nahm, ``The politics of technology bans: Industrial policy
  competition and green goals for the auto industry,'' {\em Energy Policy},
  vol.~126, pp.~470--479, 2019.

\bibitem{Roadtozero}
{UK Department of Transport}, ``The road to zero,'' 2018.

\bibitem{Ito2013}
N.~Ito, K.~Takeuchi, and S.~Managi, ``Willingness-to-pay for infrastructure
  investments for alternative fuel vehicles,'' {\em Transportation Research
  Part D: Transport and Environment}, vol.~18, pp.~1--8, 2013.

\bibitem{Hackbarth2013}
A.~Hackbarth and R.~Madlener, ``Consumer preferences for alternative fuel
  vehicles: A discrete choice analysis,'' {\em Transportation Research Part D:
  Transport and Environment}, vol.~25, pp.~5--17, 2013.

\bibitem{Abbasi2011}
T.~Abbasi and S.~Abbasi, ```renewable' hydrogen: Prospects and challenges,''
  {\em Renewable and Sustainable Energy Reviews}, vol.~15, pp.~3034--3040, aug
  2011.

\bibitem{Momirlan2005}
M.~Momirlan and T.~Veziroglu, ``The properties of hydrogen as fuel tomorrow in
  sustainable energy system for a cleaner planet,'' {\em International Journal
  of Hydrogen Energy}, vol.~30, pp.~795--802, jul 2005.

\bibitem{TolgaBalta2009}
M.~{Tolga Balta}, I.~Dincer, and A.~Hepbasli, ``Thermodynamic assessment of
  geothermal energy use in hydrogen production,'' {\em International Journal of
  Hydrogen Energy}, vol.~34, pp.~2925--2939, apr 2009.

\bibitem{Fayaz2012}
H.~Fayaz, R.~Saidur, N.~Razali, F.~Anuar, A.~Saleman, and M.~Islam, ``An
  overview of hydrogen as a vehicle fuel,'' {\em Renewable and Sustainable
  Energy Reviews}, vol.~16, pp.~5511--5528, oct 2012.

\bibitem{Bolat2014}
P.~Bolat and C.~Thiel, ``Hydrogen supply chain architecture for bottom-up
  energy systems models. part 1: Developing pathways,'' {\em International
  Journal of Hydrogen Energy}, vol.~39, pp.~8881--8897, jun 2014.

\bibitem{Dincer2015}
I.~Dincer and C.~Acar, ``Review and evaluation of hydrogen production methods
  for better sustainability,'' {\em International Journal of Hydrogen Energy},
  vol.~40, pp.~11094--11111, sep 2015.

\bibitem{Liu2012}
H.~Liu, A.~Almansoori, M.~Fowler, and A.~Elkamel, ``Analysis of ontario's
  hydrogen economy demands from hydrogen fuel cell vehicles,'' {\em
  International Journal of Hydrogen Energy}, vol.~37, pp.~8905--8916, jun 2012.

\bibitem{Dagdougui2012}
H.~Dagdougui, A.~Ouammi, and R.~Sacile, ``Modelling and control of hydrogen and
  energy flows in a network of green hydrogen refuelling stations powered by
  mixed renewable energy systems,'' {\em International Journal of Hydrogen
  Energy}, vol.~37, pp.~5360--5371, mar 2012.

\bibitem{Edwards2008}
P.~Edwards, V.~Kuznetsov, W.~David, and N.~Brandon, ``Hydrogen and fuel cells:
  Towards a sustainable energy future,'' {\em Energy Policy}, vol.~36,
  pp.~4356--4362, dec 2008.

\bibitem{Neef2009}
H.-J. Neef, ``International overview of hydrogen and fuel cell research,'' {\em
  Energy}, vol.~34, pp.~327--333, mar 2009.

\bibitem{Mazloomi2012}
K.~Mazloomi and C.~Gomes, ``Hydrogen as an energy carrier: Prospects and
  challenges,'' {\em Renewable and Sustainable Energy Reviews}, vol.~16,
  pp.~3024--3033, jun 2012.

\bibitem{Pudukudy2014}
M.~Pudukudy, Z.~Yaakob, M.~Mohammad, B.~Narayanan, and K.~Sopian, ``Renewable
  hydrogen economy in asia - opportunities and challenges: An overview,'' {\em
  Renewable and Sustainable Energy Reviews}, vol.~30, pp.~743--757, feb 2014.

\bibitem{Sharma2015}
S.~Sharma and S.~K. Ghoshal, ``Hydrogen the future transportation fuel: From
  production to applications,'' {\em Renewable and Sustainable Energy Reviews},
  vol.~43, pp.~1151--1158, mar 2015.

\bibitem{Almansoori2006}
A.~Almansoori and N.~Shah, ``Design and operation of a future hydrogen supply
  chain,'' {\em Chemical Engineering Research and Design}, vol.~84,
  pp.~423--438, jun 2006.

\bibitem{Stephan2004}
C.~Stephan and J.~Sullivan, ``An agent-based hydrogen vehicle/infrastructure
  model,'' in {\em Proceedings of the 2004 Congress on Evolutionary Computation
  (IEEE Cat. No. 04TH8753)}, vol.~2, pp.~1774--1779, IEEE, 2004.

\bibitem{Brand2020}
C.~Brand, J.~Anable, I.~Ketsopoulou, and J.~Watson, ``Road to zero or road to
  nowhere? disrupting transport and energy in a zero carbon world,'' {\em
  Energy Policy}, vol.~139, p.~111334, 2020.

\bibitem{EU2020}
{European Commission}, ``A hydrogen strategy for a climate-neutral europe,''
  2020.
\newblock
  \newline\url{https://ec.europa.eu/energy/sites/ener/files/hydrogen_strategy.pdf}.

\bibitem{METI2019}
{Ministry of Economy, Trade and Industry of Japan}, ``New strategic roadmap for
  hydrogen and fuel cells,'' 2019.
\newblock
  \newline\url{https://www.meti.go.jp/english/press/2019/0918_001.html}.

\bibitem{Kloess2011}
M.~Kloess and A.~M{\"u}ller, ``Simulating the impact of policy, energy prices
  and technological progress on the passenger car fleet in austria—a model
  based analysis 2010--2050,'' {\em Energy Policy}, vol.~39, no.~9,
  pp.~5045--5062, 2011.

\bibitem{Wang2011}
G.~Wang, ``Advanced vehicles: costs, energy use, and macroeconomic impacts,''
  {\em Journal of Power Sources}, vol.~196, no.~1, pp.~530--540, 2011.

\bibitem{Park2011}
S.~Y. Park, J.~W. Kim, and D.~H. Lee, ``Development of a market penetration
  forecasting model for hydrogen fuel cell vehicles considering infrastructure
  and cost reduction effects,'' {\em Energy Policy}, vol.~39, pp.~3307--3315,
  jun 2011.

\bibitem{Moreno2017}
M.~Moreno-Benito, P.~Agnolucci, and L.~G. Papageorgiou, ``Towards a sustainable
  hydrogen economy: Optimisation-based framework for hydrogen infrastructure
  development,'' {\em Computers \& Chemical Engineering}, vol.~102,
  pp.~110--127, 2017.

\bibitem{Samsatli2019}
S.~Samsatli and N.~J. Samsatli, ``The role of renewable hydrogen and
  inter-seasonal storage in decarbonising heat--comprehensive optimisation of
  future renewable energy value chains,'' {\em Applied Energy}, vol.~233,
  pp.~854--893, 2019.

\bibitem{Zhao2006}
J.~Zhao and M.~W. Melaina, ``Transition to hydrogen-based transportation in
  china: lessons learned from alternative fuel vehicle programs in the united
  states and china,'' {\em Energy Policy}, vol.~34, no.~11, pp.~1299--1309,
  2006.

\bibitem{Shafiei2017}
E.~Shafiei, B.~Davidsdottir, J.~Leaver, H.~Stefansson, and E.~I. Asgeirsson,
  ``Energy, economic, and mitigation cost implications of transition toward a
  carbon-neutral transport sector: A simulation-based comparison between
  hydrogen and electricity,'' {\em Journal of cleaner production}, vol.~141,
  pp.~237--247, 2017.

\bibitem{Musti2011}
S.~Musti and K.~M. Kockelman, ``Evolution of the household vehicle fleet:
  Anticipating fleet composition, phev adoption and ghg emissions in austin,
  texas,'' {\em Transportation Research Part A: Policy and Practice}, vol.~45,
  no.~8, pp.~707--720, 2011.

\bibitem{act2019}
{UK Government}, ``The climate change act 2008 (2050 target amendment) order
  2019,'' 2019.
\newblock \newline\url{https://www.legislation.gov.uk/ukpga/2008/27/section/1}.

\bibitem{act2017}
{UK Government}, ``Air quality plan for nitrogen dioxide (no2) in uk (2017),''
  2019.
\newblock
  \newline\url{https://www.gov.uk/government/publications/air-quality-plan-for-nitrogen-dioxide-no2-in-uk-2017}.

\bibitem{Morrison2012}
F.~Morrison, {\em The art of modeling dynamic systems: forecasting for chaos,
  randomness and determinism}.
\newblock Courier Corporation, 2012.

\bibitem{Wang2016}
H.-T. Wang and T.-C. Wang, ``Application of the grey lotka{\textendash}volterra
  model to forecast the diffusion and competition analysis of the {TV} and
  smartphone industries,'' {\em Technological Forecasting and Social Change},
  vol.~106, pp.~37--44, may 2016.

\bibitem{Zhang2014}
H.~Zhang, Z.~Ma, G.~Xie, and L.~Jia, ``Effects of behavioral tactics of
  predators on dynamics of a predator-prey system,'' {\em Advances in
  Mathematical Physics}, vol.~2014, pp.~1--10, 2014.

\bibitem{Gokmen2015}
E.~Gokmen, O.~R. Isik, and M.~Sezer, ``Taylor collocation approach for delayed
  lotka{\textendash}volterra predator{\textendash}prey system,'' {\em Applied
  Mathematics and Computation}, vol.~268, pp.~671--684, oct 2015.

\bibitem{Evans1999}
C.~Evans and G.~Findley, ``Analytic solutions to a family of lotka--volterra
  related differential equations,'' {\em Journal of mathematical chemistry},
  vol.~25, no.~2-3, pp.~181--189, 1999.

\bibitem{Huang2001}
X.~Huang and Y.~Zu, ``The {LES} population model: essentials and relationship
  to the lotka{\textendash}volterra model,'' {\em Ecological Modelling},
  vol.~143, pp.~215--225, sep 2001.

\bibitem{Leibling2008}
D.~Leibling, ``Car ownership in great britain,'' {\em London: RAC Foundation
  for Motoring}, 2008.

\bibitem{Wei2015}
W.~Wei and L.~Cheah, ``Singapore road vehicle fleet evolution,'' in {\em
  Complex Systems Design \& Management Asia}, pp.~55--71, Springer, 2015.

\bibitem{Hugosson2016}
M.~B. Hugosson, S.~Algers, S.~Habibi, and P.~Sundbergh, ``Evaluation of the
  swedish car fleet model using recent applications,'' {\em Transport Policy},
  vol.~49, pp.~30--40, 2016.

\bibitem{Dargay2007}
J.~Dargay, D.~Gately, and M.~Sommer, ``Vehicle ownership and income growth,
  worldwide: 1960-2030,'' {\em The energy journal}, vol.~28, no.~4, 2007.

\bibitem{Rota2016}
M.~F. Rota, J.~M. Carcedo, and J.~P. Garc{\'\i}a, ``Dual approach for modelling
  demand saturation levels in the automobile market. the gompertz curve: macro
  versus micro data,'' {\em Investigaci{\'o}n econ{\'o}mica}, vol.~75, no.~296,
  pp.~43--72, 2016.

\bibitem{Campinez2018}
S.~Camp{\'\i}{\~n}ez-Romero, A.~Colmenar-Santos, C.~P{\'e}rez-Molina, and
  F.~Mur-P{\'e}rez, ``A hydrogen refuelling stations infrastructure deployment
  for cities supported on fuel cell taxi roll-out,'' {\em Energy}, vol.~148,
  pp.~1018--1031, 2018.

\bibitem{Toyota2016}
Toyota, ``Toyota mirai.''

\bibitem{Hyundai2020}
Hyundai, ``Hyundai nexo.''

\bibitem{Renault2019}
Renault, ``Renault runs on hydrogen with master {Z.E.} hydrogen an kangoo
  {Z.E.} hydrogen.''

\bibitem{Spath2000}
P.~L. Spath and M.~K. Mann, ``Life cycle assessment of hydrogen production via
  natural gas steam reforming,'' tech. rep., sep 2000.

\bibitem{Nocera2016}
S.~Nocera and F.~Cavallaro, ``The competitiveness of alternative transport
  fuels for co2 emissions,'' {\em Transport Policy}, vol.~50, pp.~1--14, 2016.

\bibitem{Engstrom2019}
E.~Engstr{\"o}m, S.~Algers, and M.~B. Hugosson, ``The choice of new private and
  benefit cars vs. climate and transportation policy in sweden,'' {\em
  Transportation Research Part D: Transport and Environment}, vol.~69,
  pp.~276--292, 2019.

\bibitem{Seo2020}
S.~Seo, D.~Yun, and C.~Lee, ``Design and optimization of a hydrogen supply
  chain using a centralized storage model,'' {\em Applied Energy}, vol.~262,
  p.~114452, 2020.

\bibitem{France2018}
{French Government}, ``Hydrogen plan: ‘making our country a world leader in
  this technology,'' 2019.
\newblock
  \newline\url{https://www.gouvernement.fr/en/hydrogen-plan-making-our-country-a-world-leader-in-this-technology-0}.

\bibitem{Tlili2020}
O.~Tlili, C.~Mansilla, J.~Linsen, M.~Reuss, T.~Grube, M.~Robinius,
  J.~Andr{\'e}, Y.~Perez, A.~{Le Duigou}, and D.~Stolten, ``Geospatial
  modelling of the hydrogen infrastructure in france in order to identify the
  most suited supply chains,'' {\em International Journal of Hydrogen Energy},
  vol.~45, no.~4, pp.~3053--3072, 2020.

\bibitem{Chu2012}
B.~Chu, S.~Duncan, A.~Papachristodoulou, and C.~Hepburn, ``Using economic model
  predictive control to design sustainable policies for mitigating climate
  change,'' in {\em 2012 IEEE 51st IEEE Conference on Decision and Control
  (CDC)}, pp.~406--411, IEEE, 2012.

\end{thebibliography}

\end{document}